\documentclass{article}

\usepackage[preprint]{neurips_2026}

\usepackage[utf8]{inputenc}
\usepackage[T1]{fontenc}
\usepackage{url}
\usepackage{booktabs}
\usepackage{graphicx}
\usepackage{amsmath}
\usepackage{amsfonts}
\usepackage{amssymb}
\usepackage{bm}
\usepackage{braket}
\usepackage{nicefrac}
\usepackage{microtype}
\usepackage{todonotes}
\usepackage{graphicx}
\usepackage{subcaption}
\usepackage{float}
\usepackage{algorithm}
\usepackage{algpseudocode}
\usepackage{siunitx}
\usepackage{fvextra}
\usepackage{placeins} 

\usepackage[dvipsnames]{xcolor}         
\definecolor{navy}{rgb}{0.0, 0.0, 0.5}
\definecolor{ourgreen}{RGB}{118, 185, 0} 
\usepackage{hyperref}
\hypersetup{
    colorlinks=true,
    linkcolor=navy,
    filecolor=navy,      
    urlcolor=navy,
    citecolor=navy,
    pdftitle={Sparse Hessian AD},
}

\newcommand{\method}[1]{\texttt{#1}}

\newcommand{\Hess}{\mathbf{H}}

\title{Colour me shocked: \\Exact Molecular Hessians from local MLIPs in O(N) time using sparse differentiation!}

\author{
  Luca Thiede* \\
  Department of Computer Science\\
  University of Toronto\\
  Vector Institute for Artificial Intelligence \\
  Toronto, Canada \\
  \texttt{luca.thiede@yahoo.com} \\
  \And
  Andreas Burger* \\
  Department of Computer Science \\
  University of Toronto\\
  Vector Institute for Artificial Intelligence \\
  Toronto, Canada \\
  \AND
  Al\'an Aspuru-Guzik \\
  Department of Computer Science, University of Toronto \\
  Vector Institute for Artificial Intelligence \\
  Acceleration Consortium \\
  Senior Fellow, Canadian Institute for Advanced Research (CIFAR) \\
  NVIDIA \\
  Toronto, Canada \\
}

\begin{document}

\maketitle

\begin{abstract}
The Hessian of the energy with respect to the nuclear positions is indispensable in atomistic modelling. However, constructing this matrix requires $O(N)$ Hessian vector products, traditionally limiting high-accuracy Hessians to small systems. Machine learning interatomic potentials (MLIPs) have accelerated atomistic modelling by providing highly accurate energies and forces at $O(N)$ cost, yet the resulting $O(N^2)$ cost of Hessians remains a practical bottleneck for large systems. 
Based on the insight that we can derive the sparsity pattern for an MLIP's Hessians in closed form, we show in this paper how to use techniques from sparse automatic differentiation to reduce the cost of a local MLIP's Hessians to a system-size-independent number of Hessian-vector products, yielding overall $O(N)$ total cost without any approximations. We benchmark our approach on a variety of systems ranging from alkane chains to water clusters to A$\beta$40 conformers. Depending on the MLIP configuration, we achieve the linear scaling regime already on relatively small systems, resulting in large runtime reductions between $2\times$-$15\times$ for these systems. This opens up the possibility of scaling high-accuracy MLIP Hessians to very large systems, such as proteins that were previously inaccessible. 

\end{abstract}




\section{Introduction}
The Hessian is a basic ingredient in atomistic modelling, entering, for example, vibrational analysis, normal-mode calculations, transition-state characterization, and second-order geometry optimization. Yet building the full Hessian is far more expensive than evaluating the underlying energy or forces, typically scaling by a factor of $O(N)$ worse.

Machine learning interatomic potentials built on architectures such as SchNet, MACE, eSEN, AimNet, and TensorNet and trained on large databases of electronic structure data, such as OMOL25 \cite{levine2025open}, now provide reliable surrogate potentials 
\cite{schutt2017schnet, batatia2022mace, fu2025esen,zubatyuk2019aimnet,simeon2023tensornet,ko2025matgl} with accuracy comparable to high level density functional theory (DFT) at a scaling of $O(N)$ due to the enforcement of cutoff radii. Forces are trivially available because they follow by differentiating the learned scalar energy with respect to Cartesian coordinates once, which roughly doubles the cost but introduces no additional asymptotic scaling compared to the energy. By contrast, constructing the full $3N \times 3N$ Cartesian Hessian requires $3N$ second-derivative probes, and $O(N^2)$ storage, thereby introducing an additional factor in system size in time and memory and leading to a total time complexity of $O(N^2)$, see section \ref{sec:background}.

Prior work in chemistry has repeatedly exploited locality and effective sparsity to reduce the cost of second derivatives. Representative examples include reduced-scaling analytic Hessians based on localized density-matrix response, sparse numerical Hessians that evaluate only important couplings, block- or fragment-reduced vibrational models for large systems, or by using Hessian information from a cheaper method to accelerate the Hessian computation of a more expensive method \cite{Kussmann2015ReducedScalingHessian,Yang2024EfficientSparseHessianVibFreq,Ghysels2011MobileBlockHessianQMMM,NakataFedorov2021FMOSecondDerivativesChapter, sanders2015compressed}. A different route is taken in recent work that achieves $O(N)$ Hessian prediction by learning second-order information directly, rather than by differentiating through the force field \cite{burger2025shoot}. However, all these methods have in common that they are only approximately correct. 
In parallel, the field of automatic differentiation developed methods to exploit sparsity in derivatives for arbitrary black-box computer programs. In contrast to the previously listed approaches, sparse automatic differentiation is guaranteed to be exact \cite{Curtis1974SparseJacobian,Newsam1983EstimationSparseJacobian,Coleman1984SparseHessianColoring,powell1979estimation,coleman1986cyclic,gebremedhin2005what}.  
The central practical limitation of sparse automatic differentiation is that it relies on knowing the sparsity pattern of the Hessian \cite{hill2025illustrated}. In general-purpose settings, obtaining that pattern requires a sparsity-detection pass, which can be costly. Here we show, however, that for a broad class of modern local message-passing interatomic potentials, the Hessian sparsity pattern follows trivially from how information propagates through a finite number of message-passing layers within a radius of neighbours. 

Once that pattern is known, the classical sparse-derivative pipeline constructs a seed matrix from an admissible colouring, evaluates compressed Hessian-vector products, and recovers the full Hessian exactly. This graph-derived sparsity removes the sparsity pattern detection pass and makes exact sparse Hessians efficient even for single Hessian evaluations of relatively small systems. By defining heuristic sparsity patterns based on a cutoff radius, we can further extend the idea more broadly to any ab-initio method as long as the Hessian is well behaved and we accept an approximation error. This leads to an efficient black-box sparse Hessian algorithm without the need to change any of the underlying quantum chemistry code.

We summarize the contributions of this paper as follows: 
\begin{enumerate}
    \item We derive the sparsity pattern for Cartesian Hessians of the most common forms of MLIPs in closed form. 
    \item We introduce \texttt{ColPackPy}, an easy to use Python interface for sparse Hessian computations, including the graph-derived sparsity construction for MLIPs, colouring via \texttt{ColPack} \cite{gebremedhin2013colpack}, compressed probing, and differentiable recovery in PyTorch.
    \item We benchmark the resulting pipeline across representative molecular systems and MLIPs and quantify the resulting speedups across favourable and unfavourable compression regimes.
\end{enumerate}

\section{Background}
\label{sec:background}
The standard way to obtain second-order information in major automatic differentiation frameworks such as Jax and PyTorch is through Hessian-vector products (HVPs) \cite{pearlmutter1994fast,griewank2008evaluating,baydin2018automatic}. For a black box scalar function \(E:\mathbb{R}^{3N}\to\mathbb{R}\) (in our case this will of course be the MLIP mapping the atomic positions to the energy), evaluated at coordinates \(\mathbf R\), let \(\mathbf g(\mathbf R)=\nabla_{\mathbf R}E(\mathbf R)\). Then for any probing vector \(\mathbf v\in\mathbb{R}^{3N}\),
\begin{equation}
\Hess(\mathbf R)\mathbf v
=
\left.\frac{d}{d\epsilon}\,\mathbf g(\mathbf R+\epsilon \mathbf v)\right|_{\epsilon=0},
\end{equation}
that is, the directional derivative of the gradient, equivalently the Jacobian-vector product of \(\mathbf g\) in the direction \(\mathbf v\) \cite{pearlmutter1994fast,griewank2008evaluating}. In settings where an exact HVP operator is unavailable, the same directional probe can be approximated by the finite difference
\begin{equation}
\Hess(\mathbf R)\mathbf v
\approx
\frac{\mathbf g(\mathbf R+\epsilon \mathbf v)-\mathbf g(\mathbf R)}{\epsilon}, \label{eq:finite_difference}
\end{equation}
Importantly, AD frameworks expose these products without explicit formation of \(\Hess\)  \cite{pearlmutter1994fast,baydin2018automatic}. In fact, to get the full dense Hessian, we usually apply HVPs with the standard basis vectors \(\{\mathbf e_i\}_{i=1}^{3N}\) to recover the explicit Hessian column by column,
\begin{equation}
\Hess(\mathbf R)\mathbf e_i = \Hess_{:,i}(\mathbf R),
\qquad
\Hess_{\mathrm{explicit}} = \Hess_{\mathrm{implicit}}\,\mathbf I,
\end{equation}
In contrast, sparse-Hessian methods reduce the number of probes by replacing the identity matrix with a compressed seed matrix derived from a colouring of the sparsity pattern, as we will describe in the next section \cite{Coleman1984SparseHessianColoring,gebremedhin2005what,gebremedhin2009efficient}; see also Fig.~\ref{fig:toc}.

\section{Sparse Automatic Differentiation}
\begin{figure}[h!]
\centering
    \centering
    \includegraphics[width=1\linewidth]{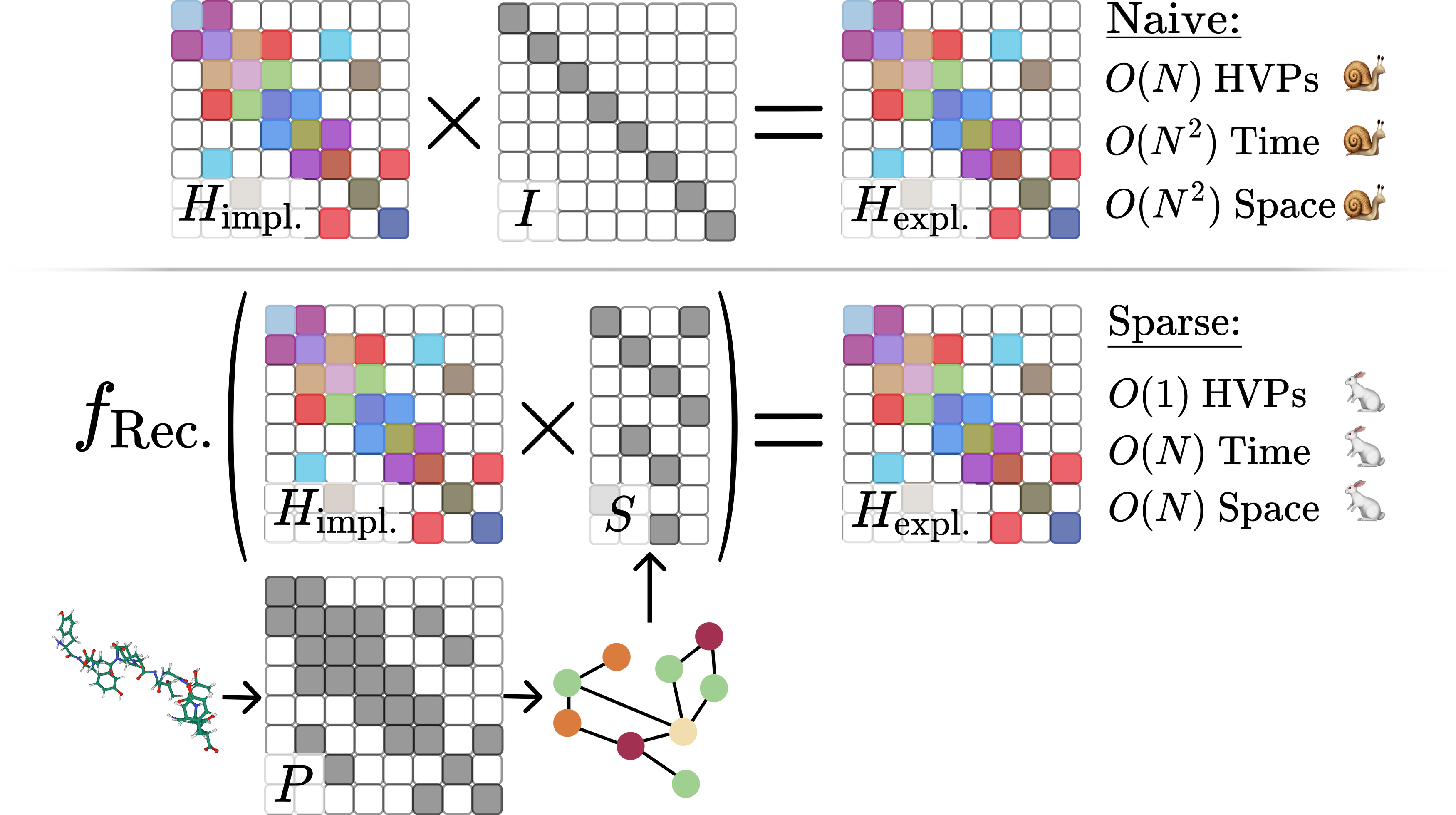}
    \caption{Top: The naive method probes each direction of the implicit Hessian individually, yielding one column of the Hessian per HVP. This requires $N$ HVPs, where $N$ is the dimension of the Hessian and can be written as the multiplication of the implicit Hessian with the identity matrix. Bottom: We first derive a sparsity pattern from the molecule and the MLIP cutoff and layer count. Using a graph colouring algorithm, this is mapped into the seed matrix $S$ which is much lower dimensional than $I$. After multiplication with the implicit Hessian $Y=H_\text{impl.}S$ we can use a recovery algorithm $f_\text{Rec.}$ to reconstruct the explicit $H_\text{expl.}$.}
\label{fig:toc}
\end{figure}
The sparse Hessian procedure employed here comprises four stages, see figure \ref{fig:toc}: 
\begin{enumerate}
    \item First we derive a sparsity pattern $\mathbf{P}$ from the atomistic graph, section \ref{sec:sparsity_pattern}
    \item We then map the sparsity pattern to a graph and use graph colouring algorithms to obtain a seed matrix $\mathbf{S} \in \{0,1\}^{3N \times p}$, whose columns are our probing vectors. Our goal is to get a seed matrix with $p$ as small as possible, while still having enough information to recover the Hessian exactly.
    \item We probe the Hessian using the seed matrix to obtain the response $\mathbf{Y} = \Hess_\text{implicit} \mathbf{S}$
    \item We use a reconstruction algorithm to get the Hessian $\Hess_\text{explicit} = f_\text{rec.}(\mathbf{Y})$
\end{enumerate}
The essential simplification relative to generic sparse-derivative settings is that we can derive the sparsity pattern $\mathbf{P}$ \emph{a priori} from the locality of the message-passing architecture (or get an approximately by a heuristic cutoff radius).

\subsection{The Sparsity Pattern of MLIPs}
\label{sec:sparsity_pattern}

To apply sparse derivative techniques, we need the sparsity pattern \(\mathbf P\) of the Hessian \cite{Coleman1984SparseHessianColoring,gebremedhin2009efficient},
\begin{align}
    P_{ij} =
    \begin{cases}
        1 & \text{if } \Hess_{ij} \neq 0,\\
        0 & \text{if } \Hess_{ij} = 0.
    \end{cases}
\end{align}
which indicates whether the Hessian entry coupling \(i\) and \(j\) can be nonzero. Detecting this sparsity pattern is often complicated and expensive \cite{gebremedhin2009efficient} for black box programs. Therefore, using sparse automatic differentiation gives the most benefits if we either reuse the same sparsity pattern across multiple calculations with different inputs, or we can use knowledge about the problem to determine the pattern efficiently. Our approach falls into the latter category.

Most modern ML interatomic potentials are local message-passing models and can be written as
\begin{align}
E_{\mathrm{tot}} = \sum_{a=1}^N E_a\bigl(\{\mathbf R_i\}_{i \in \mathcal R(a)}\bigr),
\end{align}
where \(\mathcal R(a)\) is the receptive field of atom \(a\). This form is standard for atom-centered and message-passing models for molecules and materials \cite{gilmer2017neural,schutt2017schnet,batatia2022mace}. If \(\mathbf A\) is the adjacency matrix of the interaction graph and the model consists of \(L\) message-passing layers, then
\begin{align}
\mathcal R(a) = \left\{ i \;\middle|\; (\tilde{\mathbf A}^{\,L})_{ai} \neq 0 \right\},
\qquad
\tilde{\mathbf A} = \mathbf A + \mathbf I,
\end{align}
where the identity is included so that each atom remains in its own receptive field. This is the standard graph-theoretic interpretation of \(L\) rounds of message passing: Information propagates along paths of length of at most \(L\) \cite{gilmer2017neural}. 

We define the receptive-field incidence matrix
\begin{align}
\mathbf M = \mathbf{1}\!\left[\tilde{\mathbf A}^{\,L} \neq 0\right],
\end{align}
so that \(M_{ai}=1\) exactly when atom \(i\) lies in the receptive field of center \(a\). Then a Hessian block \((i,j)\) can be nonzero only if there exists some atomic term \(E_a\) that depends on both \(\mathbf R_i\) and \(\mathbf R_j\), i.e.
\begin{align}
\boxed{P_{ij}^\text{atom}
= \mathbf{1}\!\left[\exists a \text{ such that } M_{ai} M_{aj} = 1\right]
= \mathbf{1}\!\left[(\mathbf M^\top \mathbf M)_{ij} \neq 0\right]}
\end{align}
This gives a graph-based upper bound on the atomwise Hessian sparsity pattern induced by the local decomposition of the energy \cite{schutt2017schnet,batatia2022mace}. Extensions to sparsity patterns of higher order derivatives are trivial by adding more $M_{ak}$ to the product.

So far, we have only considered nodes, but each node is embedded in Cartesian space, and therefore, the full Cartesian Hessian sparsity pattern is
\begin{align}
\boxed{\mathbf P = \mathbf P^{\mathrm{atom}} \otimes \mathbf 1_{3\times 3}}
\end{align}
Sparsity is therefore inferred from the graph topology, avoiding the computation from the sparsity-detection pass \cite{gebremedhin2009efficient}.

In some cases, the model does not permit any exact sparsity. This can occur when message-passing cutoffs are too large compared to the system size, when the network has many layers or non-local interactions such as charge equilibration, or when one seeks to apply the same sparsity logic to electronic-structure methods such as density functional theory. In these cases, we can instead use a heuristic cutoff-based sparsity pattern \(\mathbf P_{\mathrm{approx}}\). There are several reasonable choices. The simplest one is to set \(L=1\) together with a distance cutoff. However, we can also use other, more sophisticated patterns like choosing \(L>1\) and a smaller cutoff to capture long-range responses mediated by covalent bonds, or include coupling terms between formally charged groups and other prior chemical domain knowledge. A more blackbox approach is to run a cheap method like xtb \cite{grimme2017robust} or lower rung DFT with a large or no cutoff, threshold the Hessian to obtain a sparsity pattern, and use it for a more expensive method like higher rung DFT. 

Next, we use \(\mathbf P\) to construct a compressed set of probing vectors for the Hessian, so that the Hessian can be reconstructed exactly from as few directional probes as permitted by the chosen colouring and recovery scheme \cite{Coleman1984SparseHessianColoring,gebremedhin2005what,gebremedhin2009efficient}.

\subsection{From Sparsity to Seed Matrix}
\label{sec:colouring}

Given the structural sparsity pattern \(\mathbf P\) of the Hessian, we form the
associated adjacency graph
\begin{align}
    G(\mathbf P) = (V,E), \qquad
    V = \{1,\dots,n\}, \qquad
    (i,j)\in E \iff i \neq j \text{ and } P_{ij}=1.
\end{align}
This means, the vertices correspond to the columns of the sparsity pattern \(\mathbf P\) and two vertices are connected by an edge if the off-diagonal entry in sparsity pattern that connects them is non-zero.

A colouring of \(G(\mathbf P)\) partitions these columns into colour classes, which
defines a seed matrix \(\mathbf S \in \{0,1\}^{n\times p}\). If
\(c:V\to\{1,\dots,p\}\) denotes the colouring, then
\begin{align}
    S_{ik} = \mathbf 1[c(i)=k].
\end{align}
Each column of \(\mathbf S\) is therefore a probing direction that simultaneously
perturbs all variables assigned the same colour. The compressed Hessian
\begin{align}
    \mathbf Y = \mathbf H \mathbf S
\end{align}
can then be computed using one Hessian probe per colour class. The objective is to
find a colouring, and hence a seed matrix \(\mathbf S\), with as few colours as possible
while still allowing exact reconstruction of \(\mathbf H\) \cite{gebremedhin2005what,gebremedhin2009efficient}.

While there are several colouring schemes that permit exact recovery, the minimum
requirement common to all of them is that the colouring is \emph{proper}:
\begin{align}
    (i,j)\in E \;\Longrightarrow\; c(i)\neq c(j).
\end{align}
This ensures that two variables whose coupling may be nonzero are never assigned to the same probing direction. However, proper colouring alone is not sufficient for exact Hessian recovery. Even when adjacent vertices have different colours, the compressed product \(\mathbf Y\) can still mix multiple unknown Hessian entries in a way that prevents unique separation. Additional structural constraints are therefore needed to guarantee that \(\mathbf Y\) contains enough information to recover \(\Hess\) exactly. We distinguish between \emph{direct recovery}, where Hessian entries can be read off directly from \(\mathbf Y\), and \emph{substitution-based recovery}, where some entries are obtained directly from \(\mathbf Y\) and the remaining ones are recovered sequentially by subtracting contributions from entries that have already been determined \cite{gebremedhin2005what,gebremedhin2007new,gebremedhin2009efficient,montoison2025revisiting}.

For direct recovery of a symmetric Hessian, the colouring must satisfy the \method{star colouring} condition, which demands in addition to proper colouring that every path on four vertices uses at least three colours \cite{gebremedhin2007new,gebremedhin2009efficient}.

For substitution-based recovery, one can weaken this to an \method{acyclic colouring}, which demands, in addition to proper colouring, that every cycle uses at least three colours \cite{gebremedhin2007new,gebremedhin2009efficient}.

Star colouring is more restrictive, but it leads to a simpler reconstruction procedure. Acyclic colouring is less restrictive and therefore typically requires fewer colours, but the reconstruction step is more involved; in practice, indirect (substitution-based) recovery is often faster overall despite the more complicated decompression step. For implementation details on colouring and recovery, please refer to \cite{gebremedhin2009efficient}

For sparse graphs with maximum degree \(\delta\), the coloring is typically inexpensive, with commonly used heuristics scaling on the order of \(O(|V|\delta^2)\). For sparse graphs, $\delta$ becomes a constant with graph size, and the coloring itself therefore grows linear with system size.

\subsection{Scaling}
We now make the scaling regime in which the number of Hessian-vector products becomes independent of system size precise: Consider a sequence of systems with fixed MLIP cutoff radius, fixed message-passing depth $L$, and bounded local density, i.e. a bounded number of at most $d$ atoms inside any cutoff ball, independent of the total system size $N$. Then each $L$-hop receptive field has size bounded by a constant depending only on $d$ and $L$. Consequently, the induced Cartesian Hessian sparsity graph $G(P)$ has maximum degree $\delta({G(P)})$ bounded by a function dependent only on $d$ and $L$, and hence independent of the number of atoms $N$. Standard sparse-Hessian colouring results then imply that admissible star or acyclic colourings require $p = O_{d,L}(1)$
colours, corresponding to a system-size-independent number of HVPs~\cite{gebremedhin2005what}. Since each HVP through a local MLIP costs $O(N)$, the sparse Hessian can be constructed in $O(N)$ time and stored in $O(N)$ space, using a sparse tensor format.

\subsection{ColPackPy}
We package the full sparse-Hessian workflow in a lightweight Python interface using Python bindings of the C++ ColPack library \cite{gebremedhin2013colpack} for the graph-colouring algorithms, reimplementations of the recovery algorithms in PyTorch, and utility functions to derive the sparsity pattern from the connectivity graph and number of layers.
The library supports sparse return types for scalable Hessians of very large systems. For the jax ecosystem, \texttt{asdex} \cite{hill2026asdex} provides similar functionality, although no acyclic coloring is supported as of now. Concurrent work \cite{langer2026truncated} used asdex to sparsely calculate MLIP Hessian similarly to our work.

A typical usage pattern is shown below:

\begin{Verbatim}[
    numbers=left,
    breaklines,
    xleftmargin=1.5em,
    breaksymbolleft={},
    frame=lines,
    framesep=2mm,
    fontsize=\small
]
from colpackpy.torch_utils import compute_sparse_hessian_fn

def energy_fn(x_flat: torch.Tensor) -> torch.Tensor:
    pos = x_flat.view(-1, 3)
    return mlip_energy(pos)  # scalar torch.Tensor from any MLIP

sparsity_pattern, n = create_sparsity_pattern_from_graph(g, # "dgl.DGLGraph" or torch_geometric.data.Data 
    num_layers # int
    ) 

H, stats = compute_sparse_hessian_fn(
    energy_fn,
    positions, # flattened Cartesian coordinates, shape (3 * n_atoms,)
    sparsity_pattern,
    n=n,
    coloring_algorithm="ACYCLIC",   # or "STAR"
    return_stats=True,
)
\end{Verbatim}

Because the compressed probing and recovery steps are implemented in PyTorch, the resulting sparse-Hessian calculation remains differentiable. 

\section{Experiments}

The results presented here address five questions. 
\begin{enumerate}
    \item How much does sparse colouring reduce the number of HVPs?
    \item How strongly does that reduction translate into wall-clock savings?
    \item How do the savings vary across the systems studied? 
    \item Which colouring algorithm performs best?
    \item How large is the error stemming from an approximate sparsity pattern?
\end{enumerate}
 
We compare the (\method{naive}) Hessian to the sparse Hessian via direct recovery (\method{star}), and substitution-based recovery (\method{acyclic}).
All reported runtimes are timed on a Quadro RTX 5000 GPU, and all MLIP implementations and checkpoints came from the MatGL library \cite{ko2025matgl}. The maximal difference between the naive and the sparsely recovered Hessians over all elements is on the order of $10^{-8}$ for float32, so within floating point precision as expected.

\subsection{Exact recovery}
We characterize the speedup of exact sparse differentiation across three systems and two MLIP settings: We use linear alkane chains to test the best case scenario, dense water clusters \cite{Gijon2022} for the worst case, and A$\beta$40, a 40-residue protein in differently folded states for intermediate cases taken from the protein ensemble database \cite{PED00531, lazar2021ped}.
For the MLIPs, we use TensorNet \cite{simeon2023tensornet} with one layer and a cutoff of 5 \AA, similar to what is used in simulations of very large systems such as \cite{wang202529,song2024general,willman2024accuracy,liu2026smc} and TensorNet with two layers and a cutoff of 5 \AA, similar to some foundation models such as MACE-OFF \cite{kovacs2023mace}. For each system, we count the number of HVPs and the total runtime. We additionally tested MACE-OFF23 (medium) which has the same sparsity pattern as the larger tested TensorNet. Sparsity therefore saves the same number of HVPs; however, each MACE HVP is 38x more expensive than TensorNet (48ms vs 1852ms on the C300 Alkane chain on our system). The share of the total runtime of the coloring overhead therefore shrinks from 26\% for TensorNet to 0.8\% for MACE, and the speedup for MACE becomes even larger.

\begin{figure}[t]
\centering
\begin{subfigure}[t]{0.49\linewidth}
    \centering
    \includegraphics[width=\linewidth]{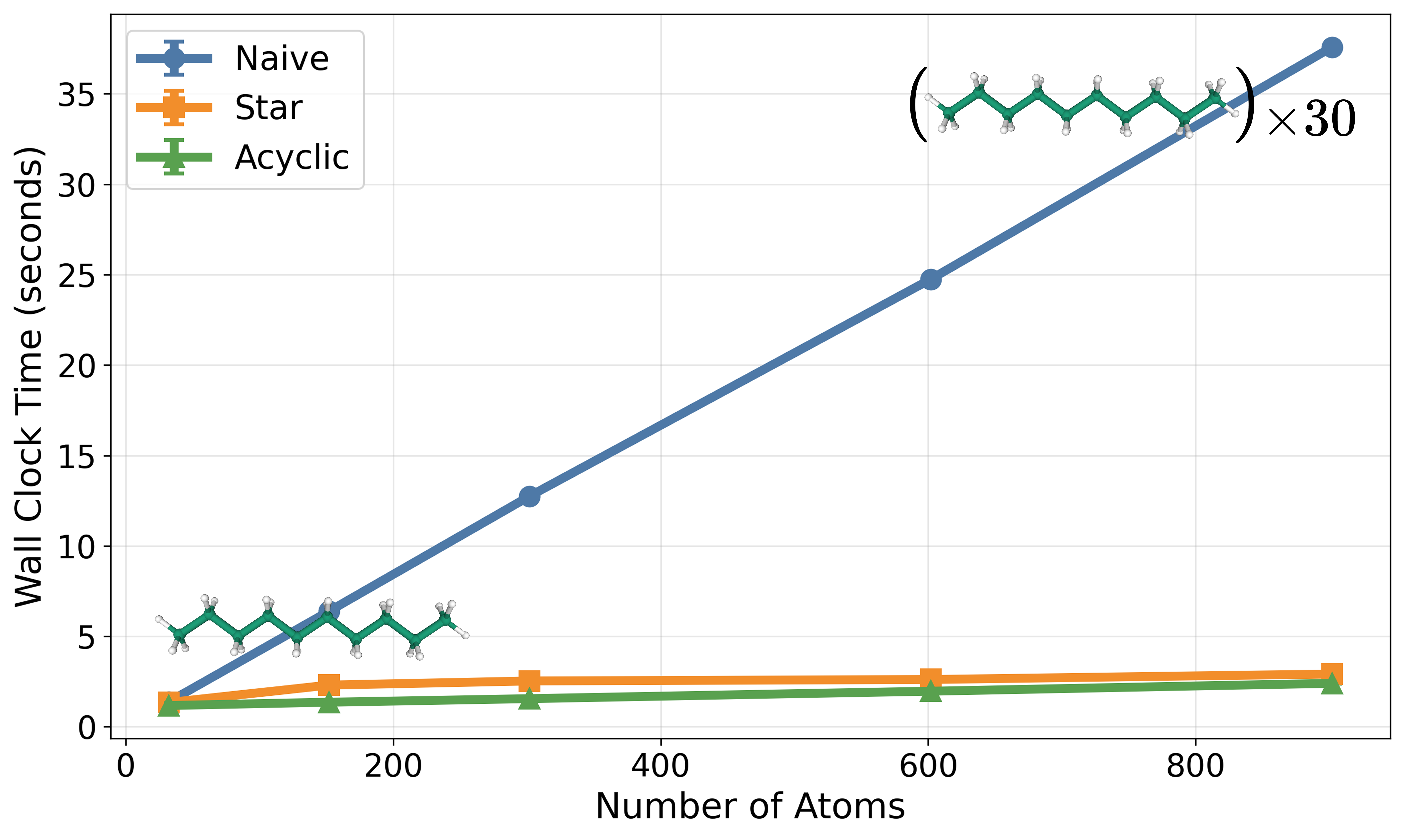}
    \caption{Runtime scaling for a one layer TensorNet on alkane chains.\label{fig:timings_alkanes_and_water_alkanes}}
\end{subfigure}\hfill
\begin{subfigure}[t]{0.49\linewidth}
    \centering
    \includegraphics[width=\linewidth]{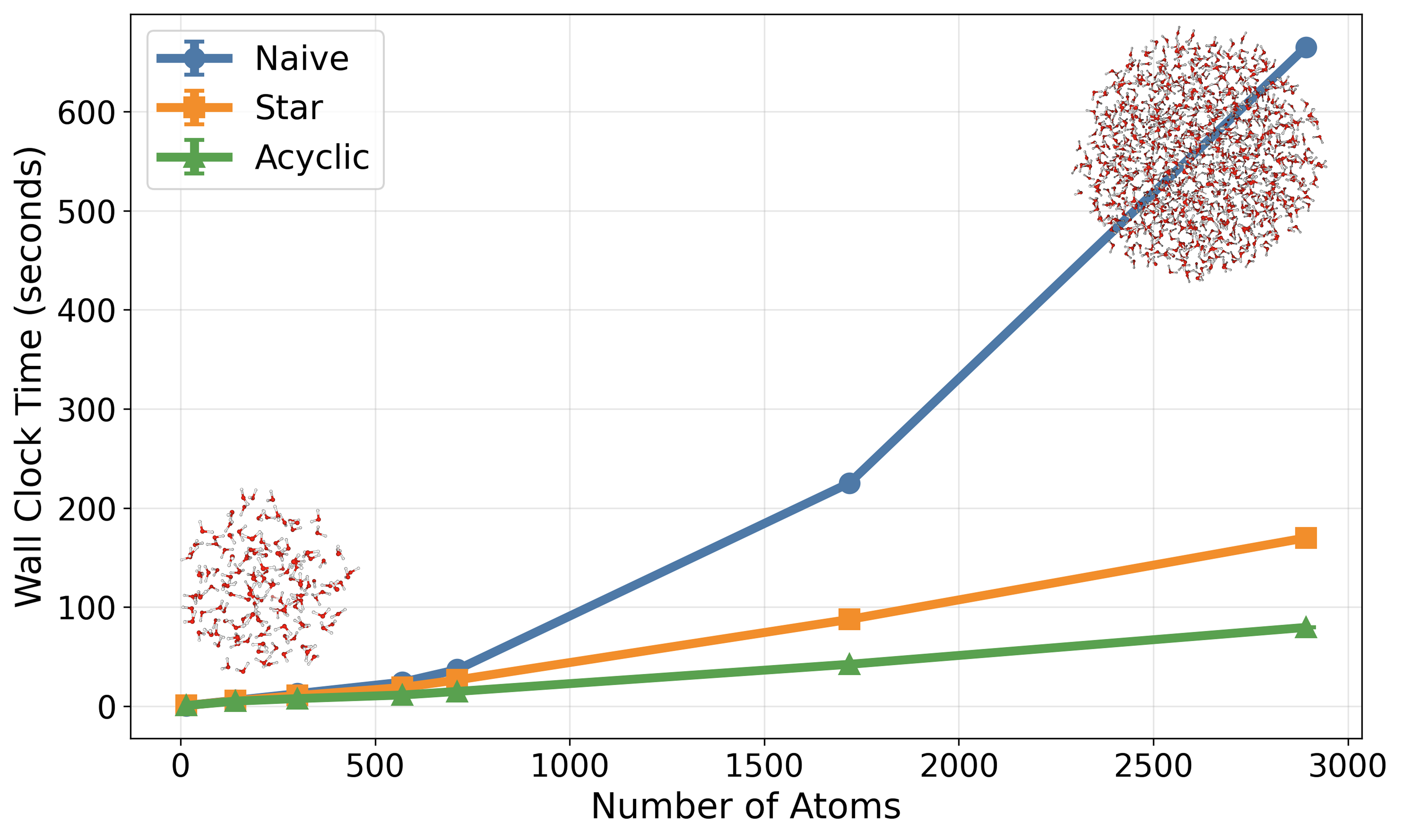}
    \caption{Runtime scaling for a one layer TensorNet on water clusters.\label{fig:timings_alkanes_and_water_water_cluster}}
\end{subfigure}
\begin{subfigure}[t]{0.49\linewidth}
    \centering
    \includegraphics[width=\linewidth]{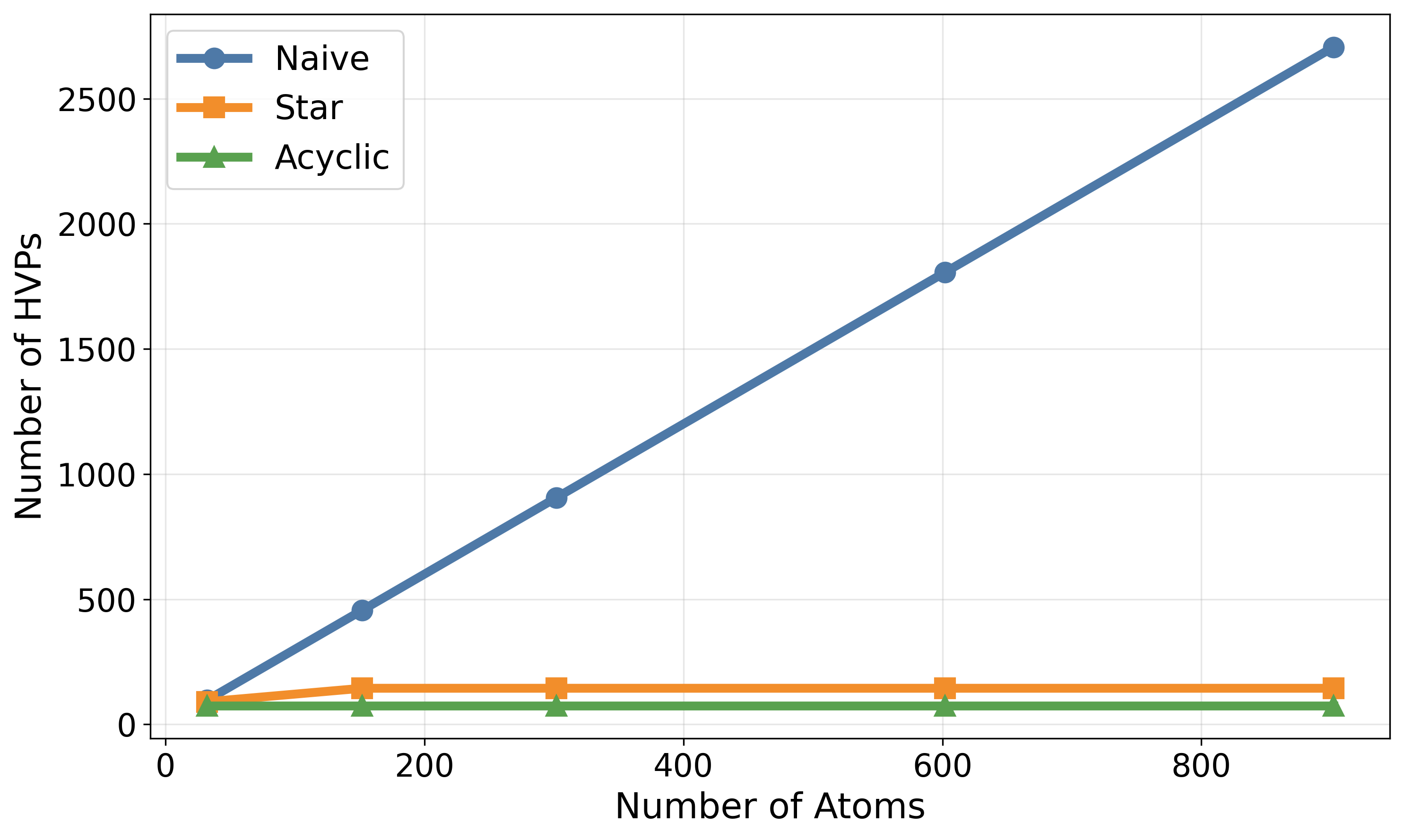}
    \caption{HVP count scaling for a one layer TensorNet on Alkane chains.}
\end{subfigure}\hfill
\begin{subfigure}[t]{0.49\linewidth}
    \centering
    \includegraphics[width=\linewidth]{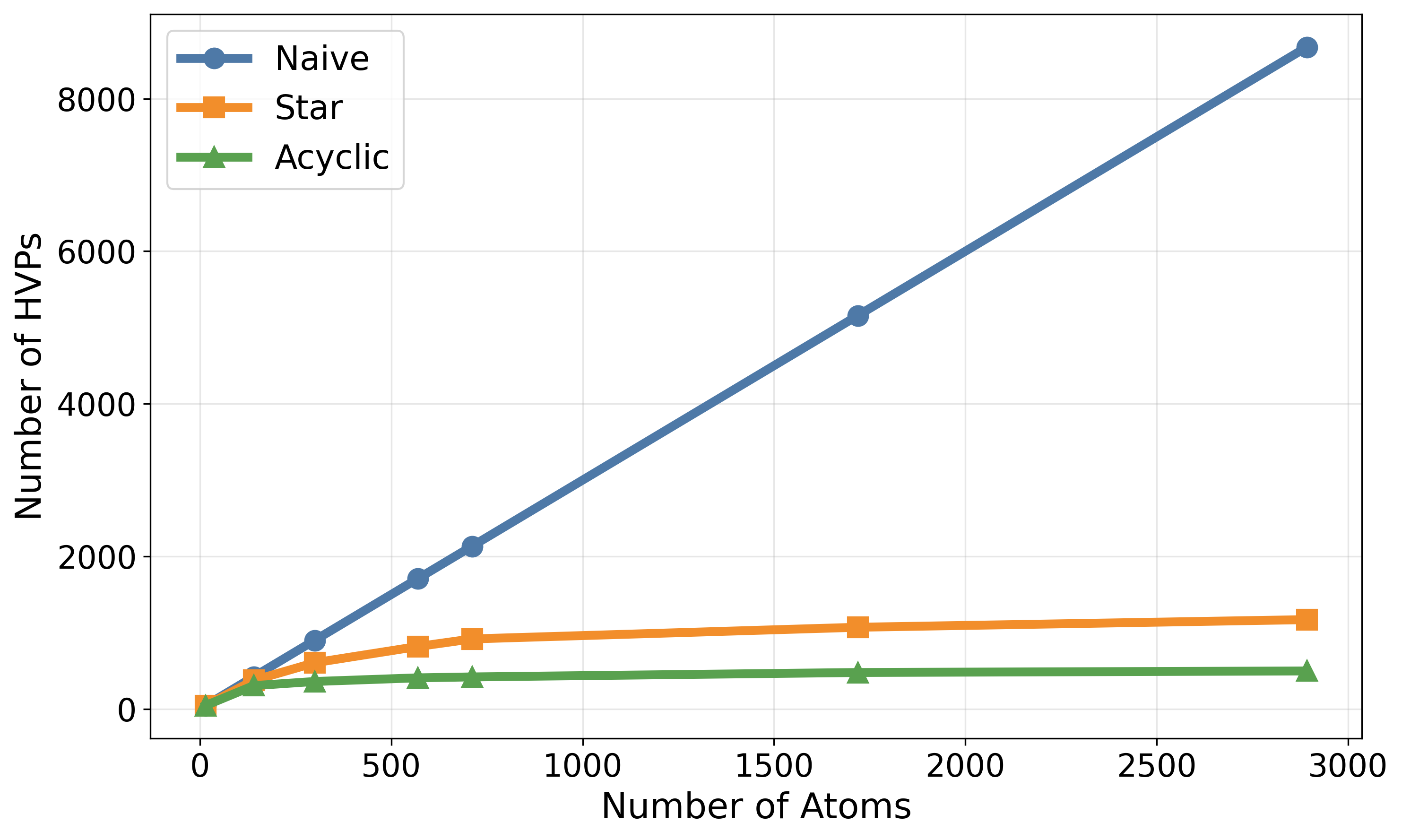}
    \caption{HVP count scaling for a one layer TensorNet on water clusters.}
\end{subfigure}
\caption{Runtime and HVP count scaling of a one layer TensorNet on increasing alkane chains and water clusters.}
\end{figure}

\paragraph{Alkanes and Water Cluster}
We plot the timings of the one-layer MLIP in \ref{fig:timings_alkanes_and_water_alkanes} and \ref{fig:timings_alkanes_and_water_water_cluster}. Plots of the HVP counts as well as the timings of the two-layer MLIPs are in appendix \ref{fig:alkanes_and_water_0l_extended} and \ref{fig:alkanes_and_water_1l_extended}. The HVP count for the \method{naive} approach grows linearly, while it tapers and becomes constant for the sparse approaches as expected. This translates into linear scaling timings even in the unfavourable dense water cluster case. Especially \method{acyclic} leads to consistent and large runtime savings of up to $15\times$ for alkanes and $8\times$ for water cluster, which is why we recommend it as the default choice. The \method{star} algorithm still reduces the HVP count but generally performs worse than the\method{acyclic} coloring and should therefore only be understood as a reference.

\paragraph{A$\mathbf{\beta}$40 Protein Conformer}
\begin{figure}[h!]
\centering
    \centering
    \includegraphics[width=\linewidth]{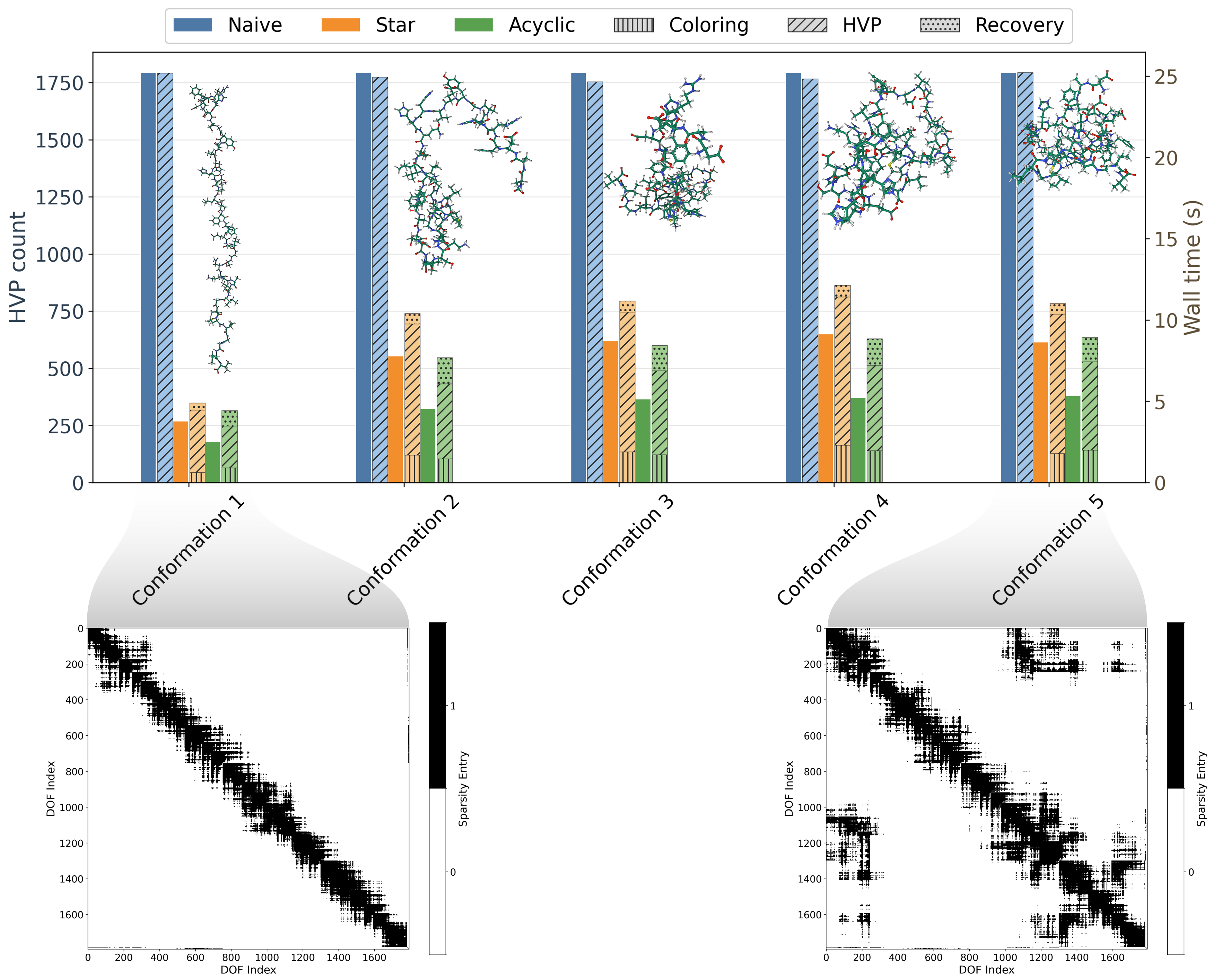}
    \caption{Runtime for computing the Hessian of a one layer TensorNet on different A$\beta$40 conformations with a runtime breakdown into the colouring, HVP and recovery step. The corresponding sparsity pattern induced by TensorNet is shown for two examples on the bottom.}
\label{fig:abeta40_one_layers}
\end{figure}
We further plot the HVP count and timings with a breakdown in colouring, HVPs, and recovery in \ref{fig:abeta40_one_layers} for the one-layer MLIP, and in the appendix \ref{fig:abeta40_two_layers} for the two-layer MLIP. We see a similar story as in the previous section, with consistent and significant savings in HVP and runtime with up to $6\times$ runtime savings.

The breakdown also reveals that the colouring, while small, is not completely negligible. The reason for this is that the HVPs are running on GPU while the colouring is currently implemented on CPU, which could be addressed in the future with GPU based colouring algorithms.

\subsection{Cutoff-based approximate recovery}
\begin{figure}[t]
\centering
\begin{subfigure}[t]{0.49\linewidth}
    \centering
\includegraphics[width=\linewidth]{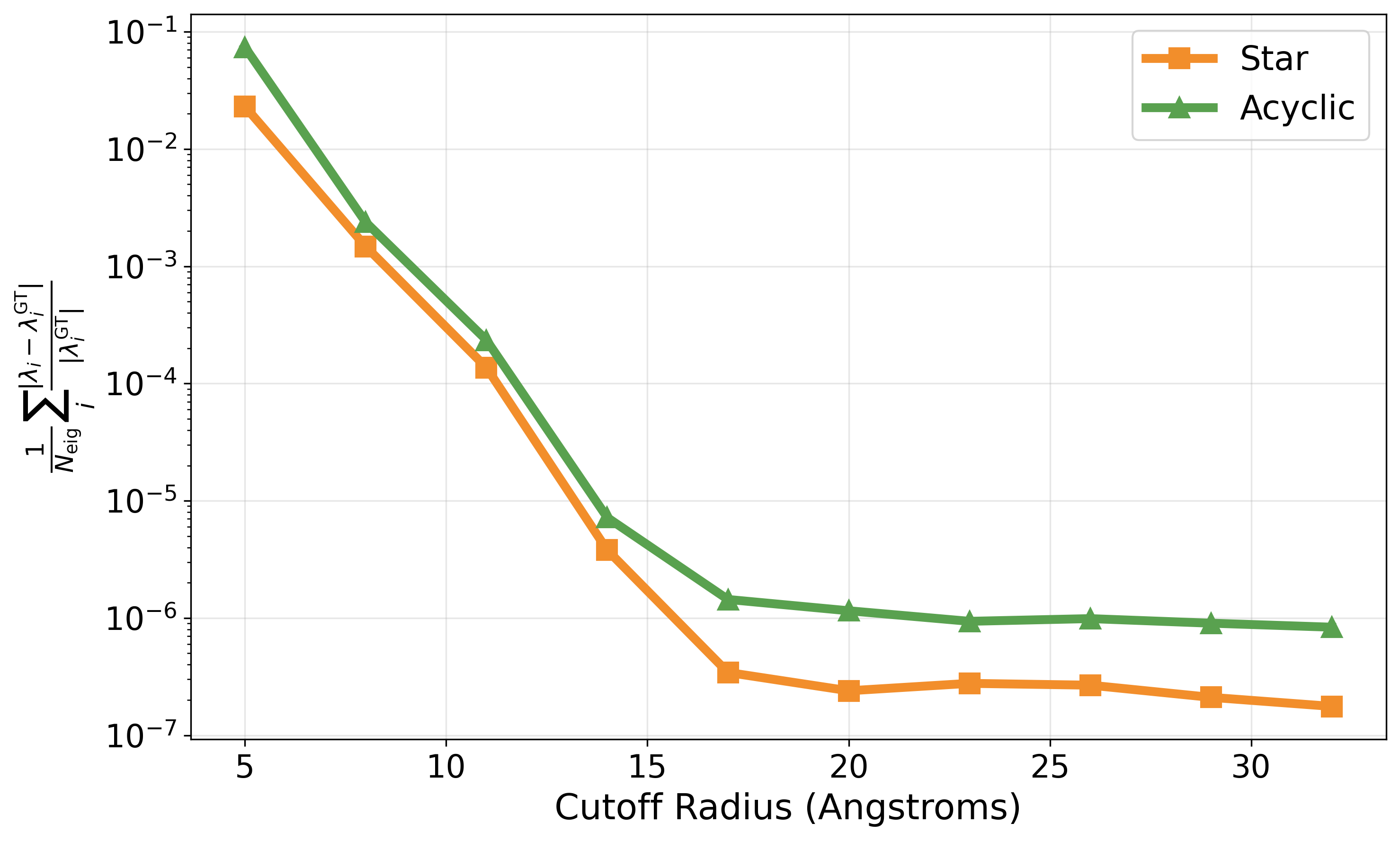}
    \caption{Average relative eigenvalue error after Eckart projection on Chignolin, dependent on the cutoff radius when used for the CHGNet MLIP with a receptive field covering the entire molecule.  \label{fig:cutoff_a}}
\end{subfigure}\hfill 
\begin{subfigure}[t]{0.49\linewidth}
    \centering
\includegraphics[width=\linewidth]{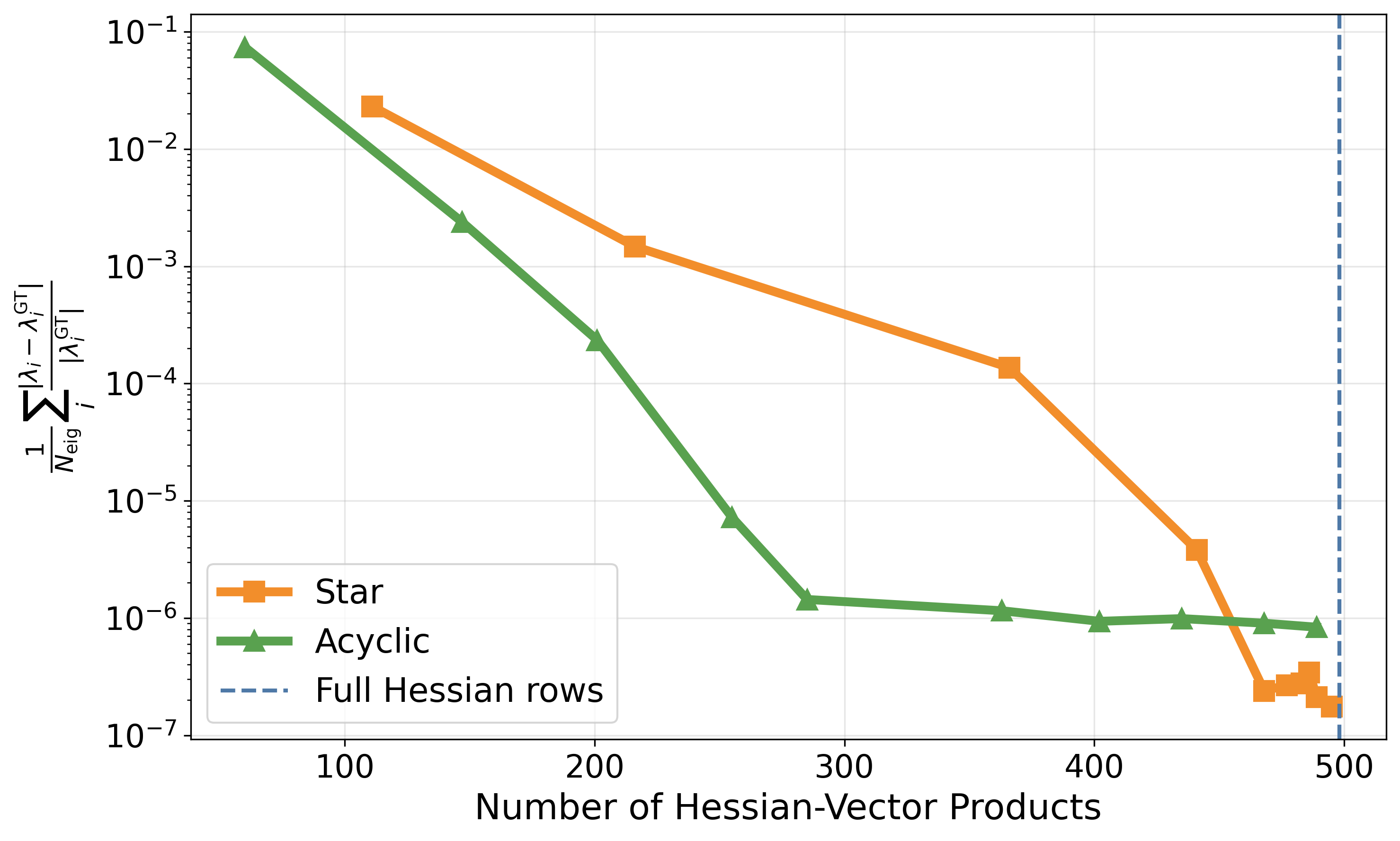}
    \caption{Average relative eigenvalue error after Eckart projection on Chignolin, dependent on the number of HVP we perform as a result of the cutoff radius implied sparsity pattern.  \label{fig:cutoff_b}}
\end{subfigure}
\caption{Approximation error induced by using an approximate cutoff-based sparsity pattern with the CHGNet MLIP with a receptive field covering the entire protein.}
\end{figure}
\begin{figure}[t]
\centering
\begin{subfigure}[t]{0.49\linewidth}
    \centering
\includegraphics[width=\linewidth]{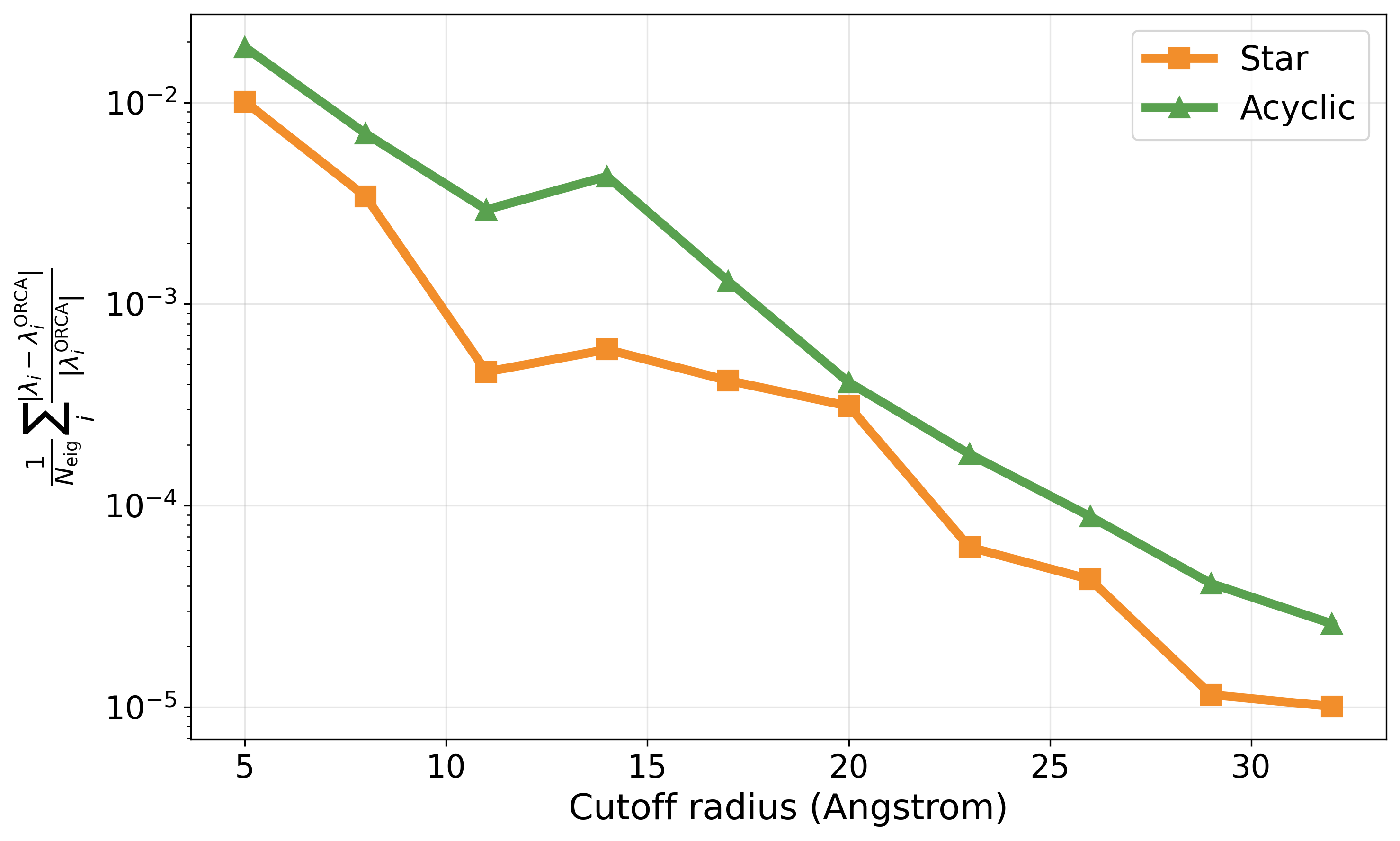}
    \caption{Average relative eigenvalue error after Eckart projection on Chignolin, dependent on the cutoff radius when used with DFT at PBE level of theory.  \label{fig:cutoff_c}}
\end{subfigure}\hfill 
\begin{subfigure}[t]{0.49\linewidth}
    \centering
\includegraphics[width=\linewidth]{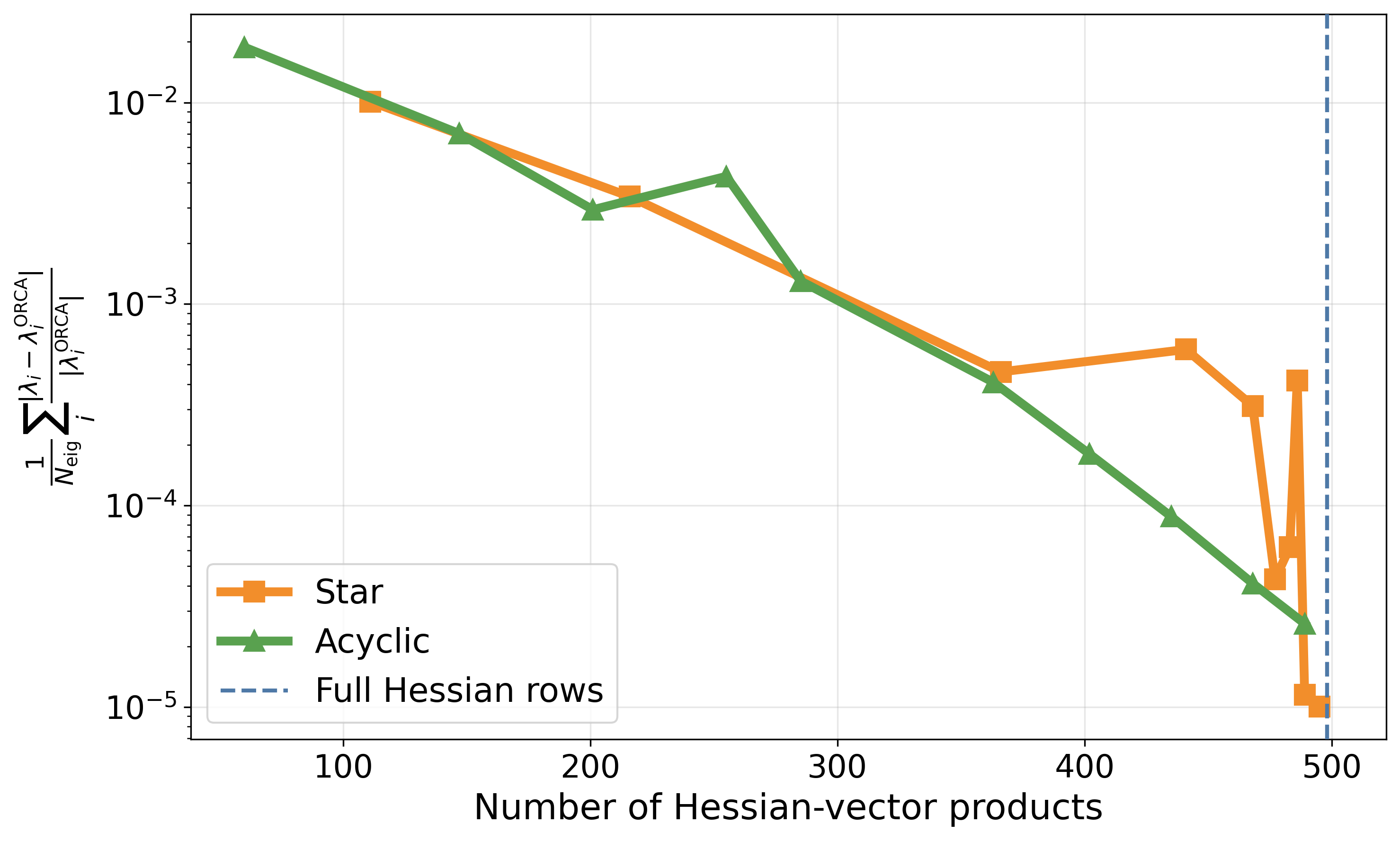}
    \caption{Average relative eigenvalue error after Eckart projection on Chignolin with DFT at PBE level of theory, dependent on the number of HVP we perform as a result of the cutoff radius implied sparsity pattern.  \label{fig:cutoff_d}}
\end{subfigure}
\caption{Approximation error induced by using an approximate cutoff-based sparsity pattern with DFT at the PBE level of theory.}
\end{figure}
In cases where the receptive field of the model becomes too large compared to the system, such that there is no exactly exploitable sparsity, we can use a heuristic sparsity pattern as discussed in section \ref{sec:sparsity_pattern}. The simplest of such heuristics is to build it from a simple atom pair cutoff radius. We test this on the Chignolin protein \cite{Wang2023} in a difficult non-equilibrium conformation with maximal span of about 32\AA, by sweeping over cutoff radii from 5 to 32 \AA and record the average relative eigenvalue error $\frac{1}{3N}\sum_i^{3N} \frac{\left|\lambda_i^\text{approx} - \lambda_i^\text{GT}\right|}{|\lambda_i^\text{GT}| + 1e^{-5}}$ after Eckart projection \cite{eckart1935some} and the number of HVPs. In the appendix \ref{sec:cutoff_details} we additionally show the maximum error over Hessian elements. We test two models:

\paragraph{CHGNet} We use a CHGNet model \cite{Deng2023CHGNet} with 5 layers and 6 \AA, resulting in a receptive field of 30 \AA, almost entirely covering the protein. The results are shown in figure \ref{fig:cutoff_a} and \ref{fig:cutoff_b}. We see that both \method{acyclic} and \method{star} get, for this system, close to numerical accuracy with an error between $10^{-6}-10^{-7}$ after about 17 \AA. However, the \method{acyclic} method clearly saves more HVPs at the 17 \AA{} cutoff. 

\paragraph{DFT} We further test the approach with DFT at the PBE level of theory on the same Chignolin protein using ORCA \cite{ORCA}. As ORCA does not expose their HVP API we emulate the HVPs. For DFT and HVP details, see appendix \ref{sec:cutoff_details}. The results are shown in figure \ref{fig:cutoff_c} and \ref{fig:cutoff_d}. Again, we drop to very low errors of about 0.1\% average eigenvalue error after about 17 \AA. However, compared to CHGNet the error drops slower, indicating that there are slightly more long-range dependencies present in DFT Hessians.

\section{Extensions and future work}
Currently, the colouring cost is usually small, but not completely negligible, as it runs on CPU. In the future, it might be possible to further reduce the runtime by adapting advances in GPU accelerated colouring algorithms to star and acyclic colouring \cite{chen2017efficient}.\\
The framework can also be extended to higher derivatives like the anharmonicity \cite{kaser2021transfer}. Our sparsity pattern derivation extends trivially to higher orders. The coloring can be achieved either by chaining lower-order derivatives or using tensor colorings \cite{deussen2019efficient}. Sparse higher-order derivatives are useful, for example, to scale anharmonic vibrational corrections like VPT2 to large systems. \\ 
Another application of our work is batched parallel computation of Hessians of different non-interacting systems. Naively, this requires $3N_\text{tot}$ HVPs with a total number of atoms $N_\text{tot}$. However, the total Hessian is block-diagonal, and using our framework, the number of HVPs can therefore be reduced to at most $3N_\text{max}$, with the maximum number of atoms over the systems in the batch $N_\text{max}$, even if each subsystem is dense.\\
Finally, it might be possible to supervise MLIP second derivatives in the compressed subspace spanned by the seed matrix, instead of random HVPs as is currently done \cite{rodriguez2026projected, koker2026pft}. 

\section{Conclusions}
By deriving the sparsity pattern for MLIPs analytically without the need for expensive detection algorithms, this paper makes sparse second-order derivatives practical, even for relatively small systems, yielding linear scaling exact Hessians while leaving the underlying force-field architecture unchanged. To make these techniques as accessible as possible, we introduce the easy to use ColPackPy library. Across various systems, the method reduces both probe counts and runtimes considerably. Real production biomolecular systems can be orders of magnitude larger than the considered systems. For example, a typical protein can have on the order of 1000 residues, compared to the 40 of our A$\beta$40 test case. Since our approach scales better, the savings for more typical systems will be even larger.

\FloatBarrier


\bibliographystyle{plainnat}
\bibliography{references}

\newpage
\appendix

\FloatBarrier
\section{Additional results and details}
We provide additional results with the 2 layer TensorNet architecture on the Alkane Chains, Water Cluster and A$\beta$40 protein. We can see that similar to the 1 layer model, we achieve linear scaling, although somewhat later as the two layer model covers more of the systems and therefore has less sparsity to exploit.
\subsection{Alkanes and Water Cluster}
\begin{figure}[ht]
\centering
\begin{subfigure}[t]{0.49\linewidth}
    \centering
    \includegraphics[width=\linewidth]{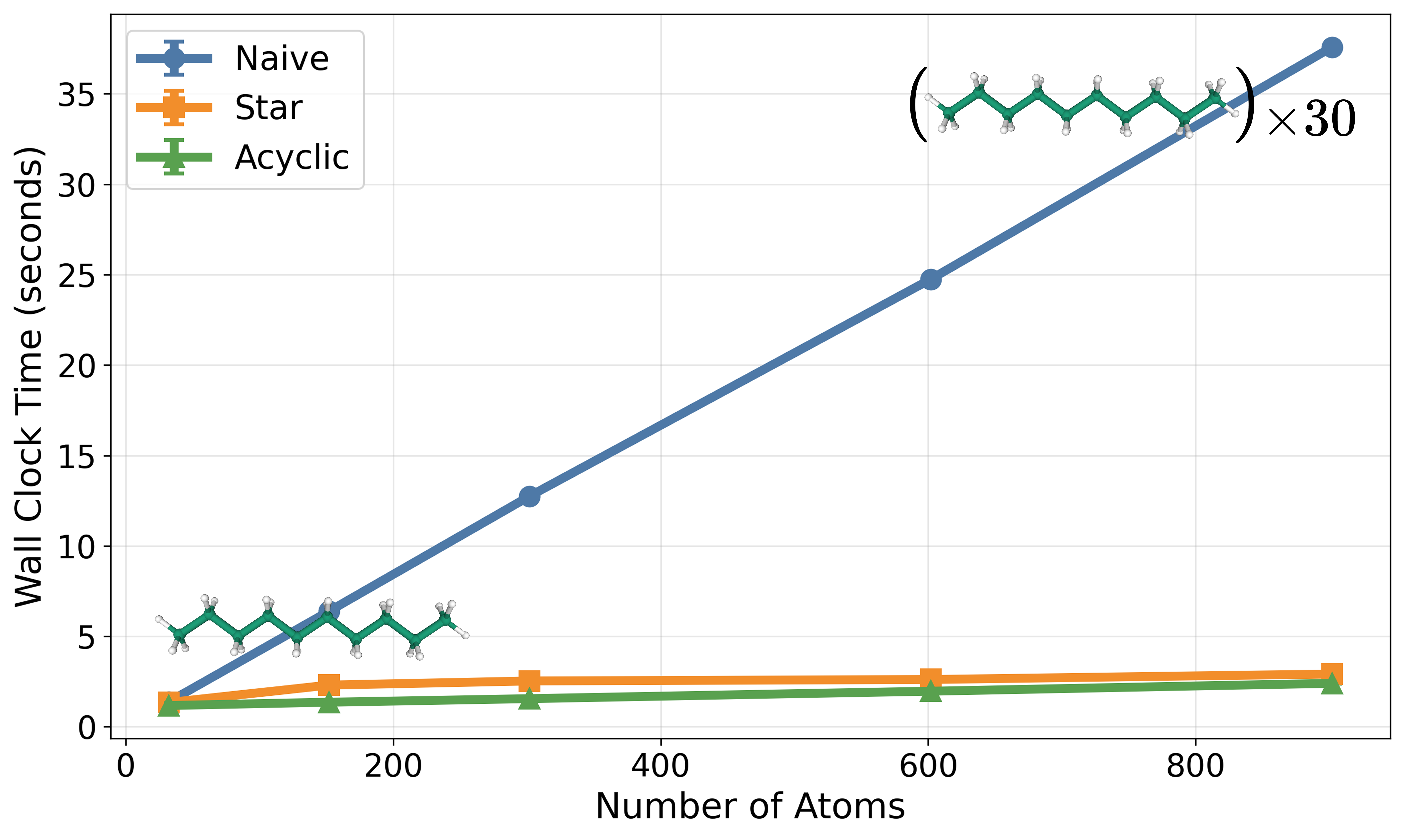}
    \caption{Alkane runtime scaling.}
\end{subfigure}\hfill
\begin{subfigure}[t]{0.49\linewidth}
    \centering
    \includegraphics[width=\linewidth]{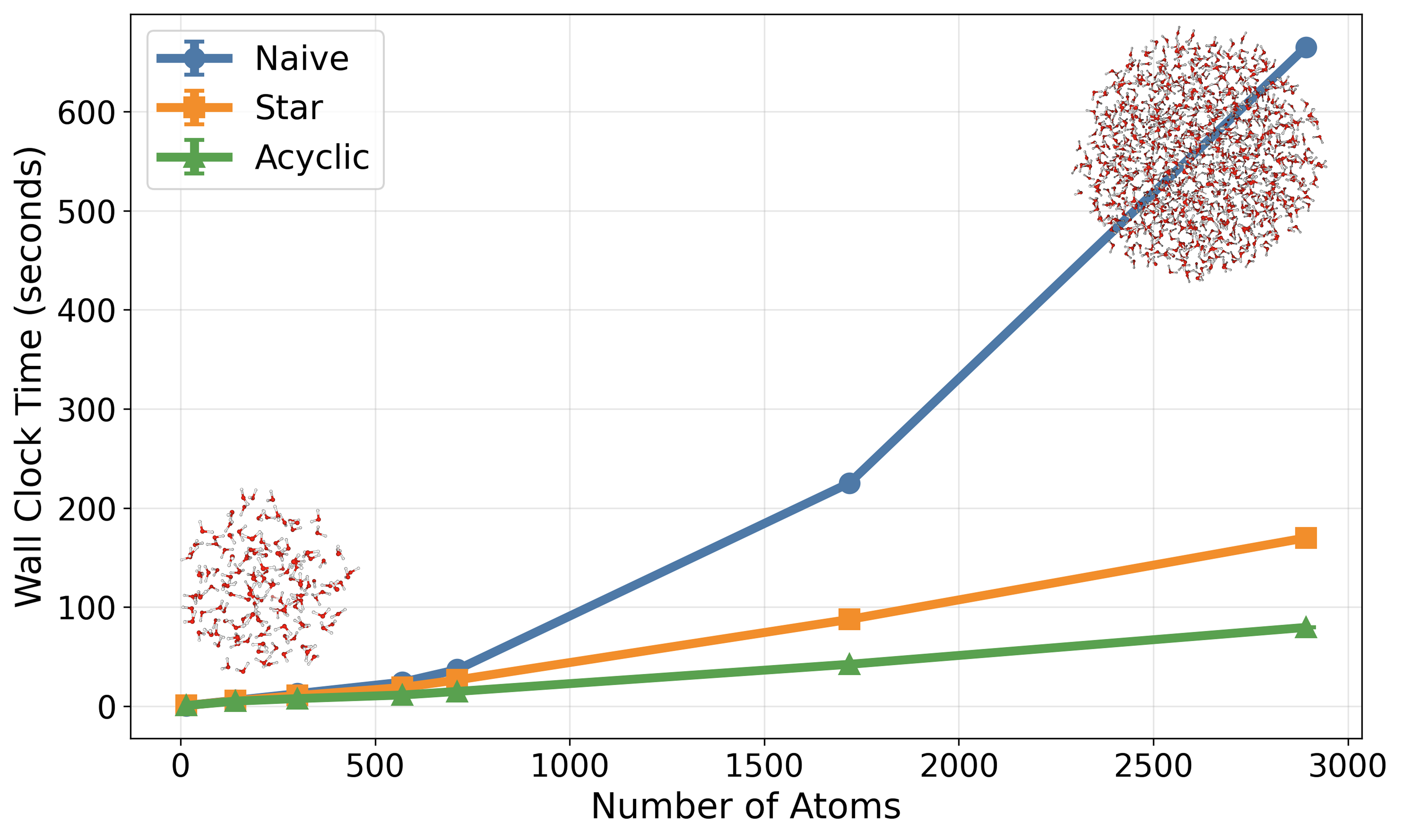}
    \caption{Water cluster runtime scaling.}
\end{subfigure}
\begin{subfigure}[t]{0.49\linewidth}
    \centering
    \includegraphics[width=\linewidth]{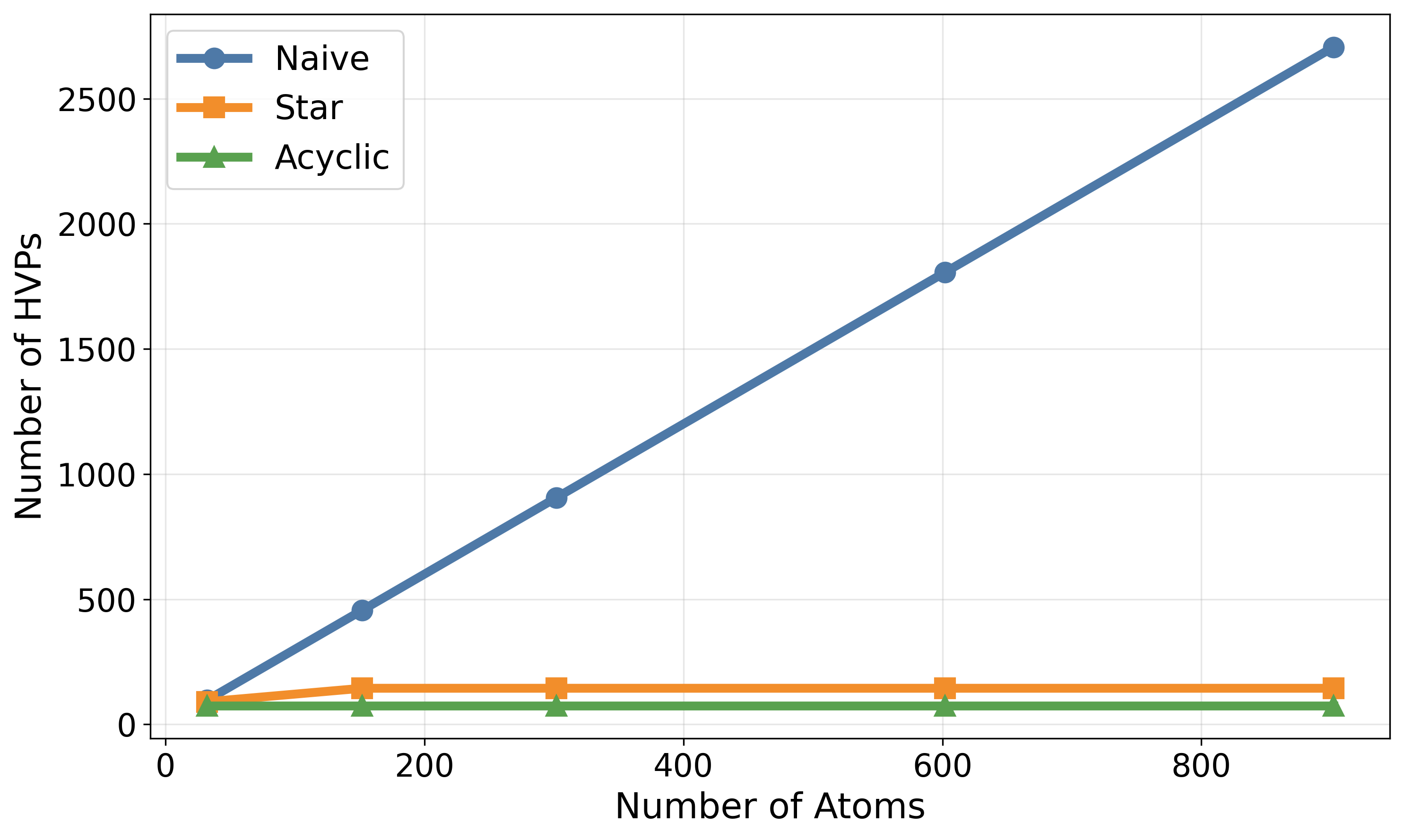}
    \caption{Alkane HVP count scaling.}
\end{subfigure}\hfill
\begin{subfigure}[t]{0.49\linewidth}
    \centering
    \includegraphics[width=\linewidth]{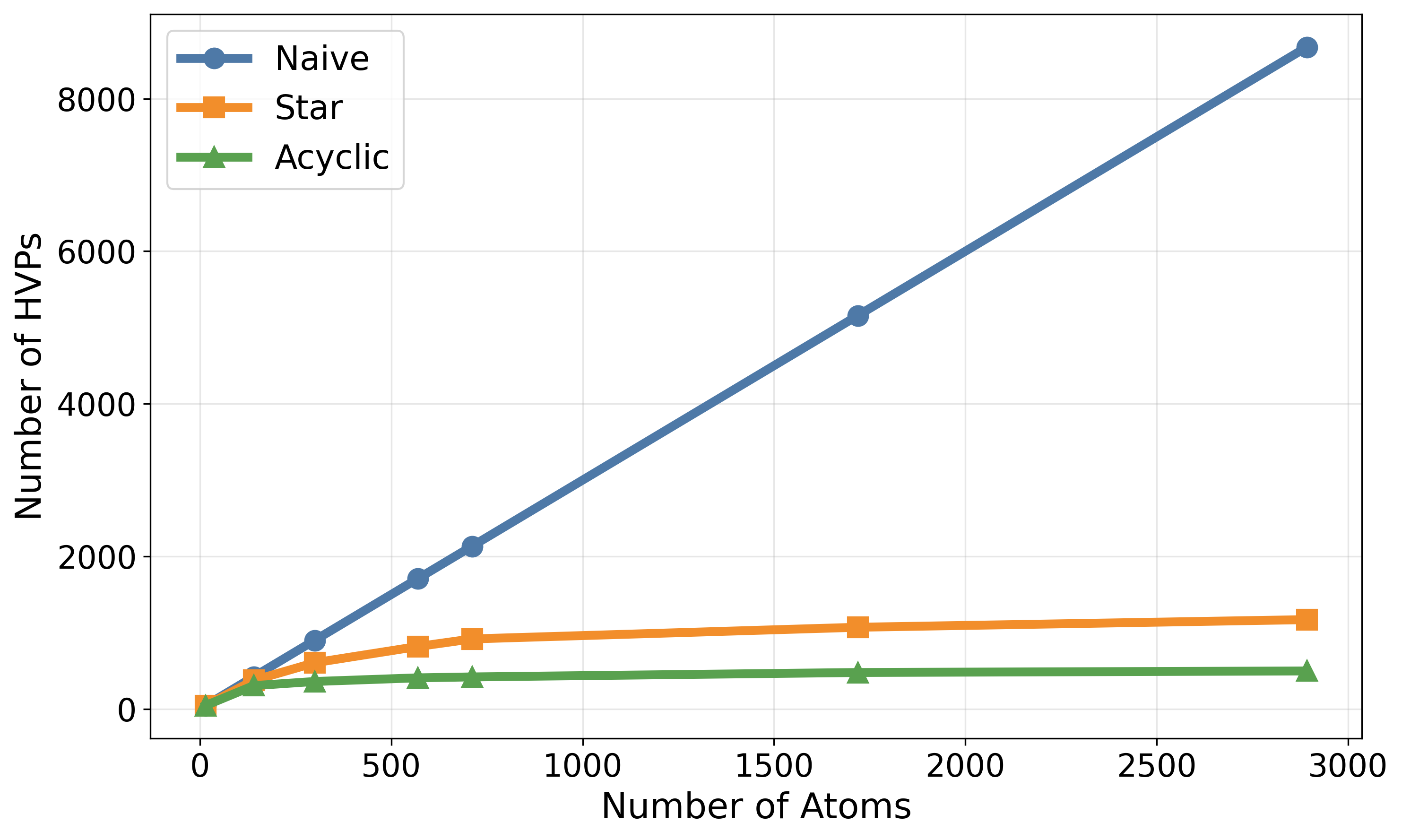}
    \caption{Water cluster HVP count scaling.}
\end{subfigure}
\caption{The runtime and HVP count of a 1 Layer TensorNet for increasingly larger alkane chains and water clusters. \label{fig:alkanes_and_water_0l_extended}}

\end{figure}

\begin{figure}[ht]
\centering
\begin{subfigure}[t]{0.49\linewidth}
    \centering
    \includegraphics[width=\linewidth]{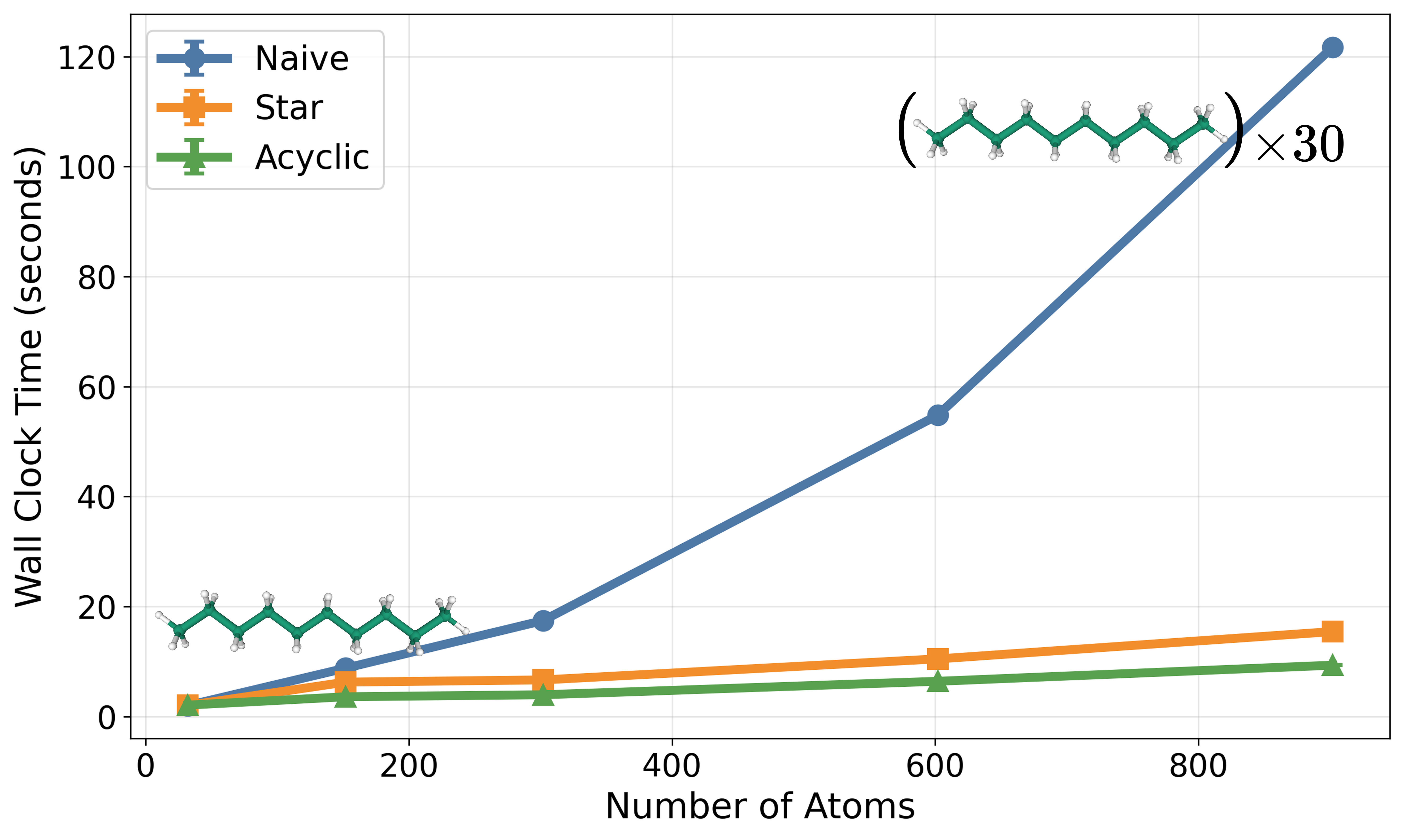}
    \caption{Alkane runtime scaling.}
\end{subfigure}\hfill
\begin{subfigure}[t]{0.49\linewidth}
    \centering
    \includegraphics[width=\linewidth]{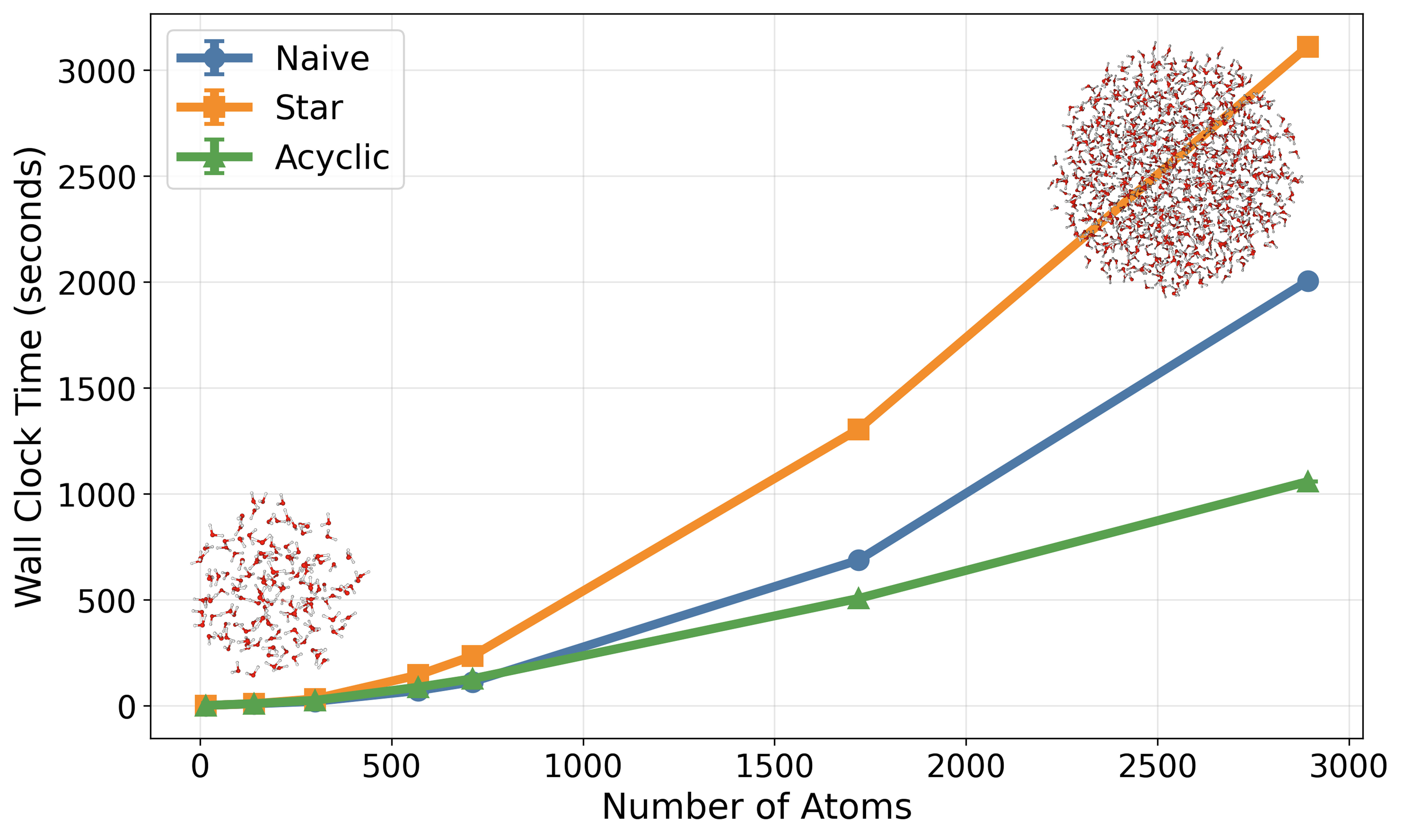}
    \caption{Water cluster runtime scaling.}
\end{subfigure}
\begin{subfigure}[t]{0.49\linewidth}
    \centering
    \includegraphics[width=\linewidth]{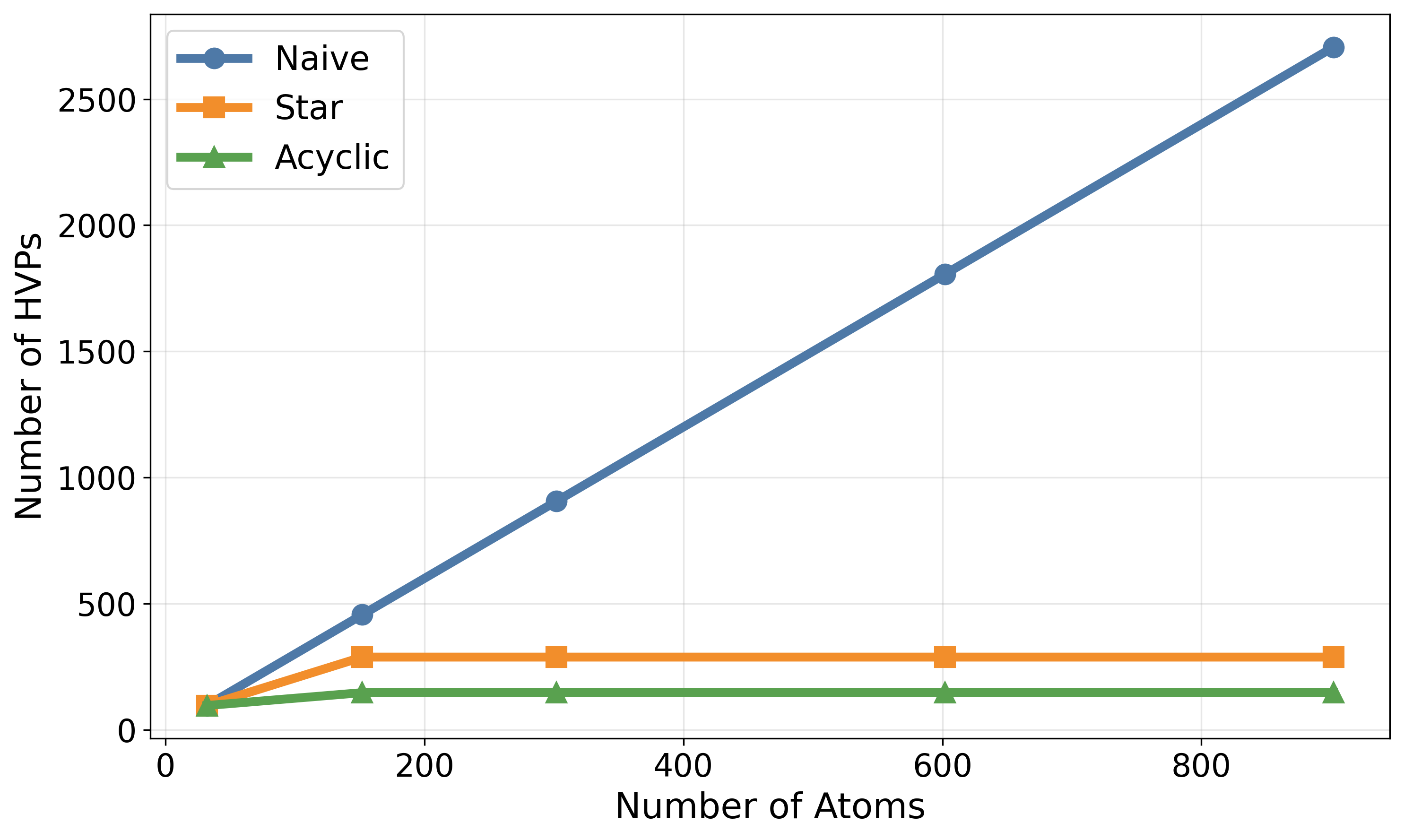}
    \caption{Alkane HVP count scaling.}
\end{subfigure}\hfill
\begin{subfigure}[t]{0.49\linewidth}
    \centering
    \includegraphics[width=\linewidth]{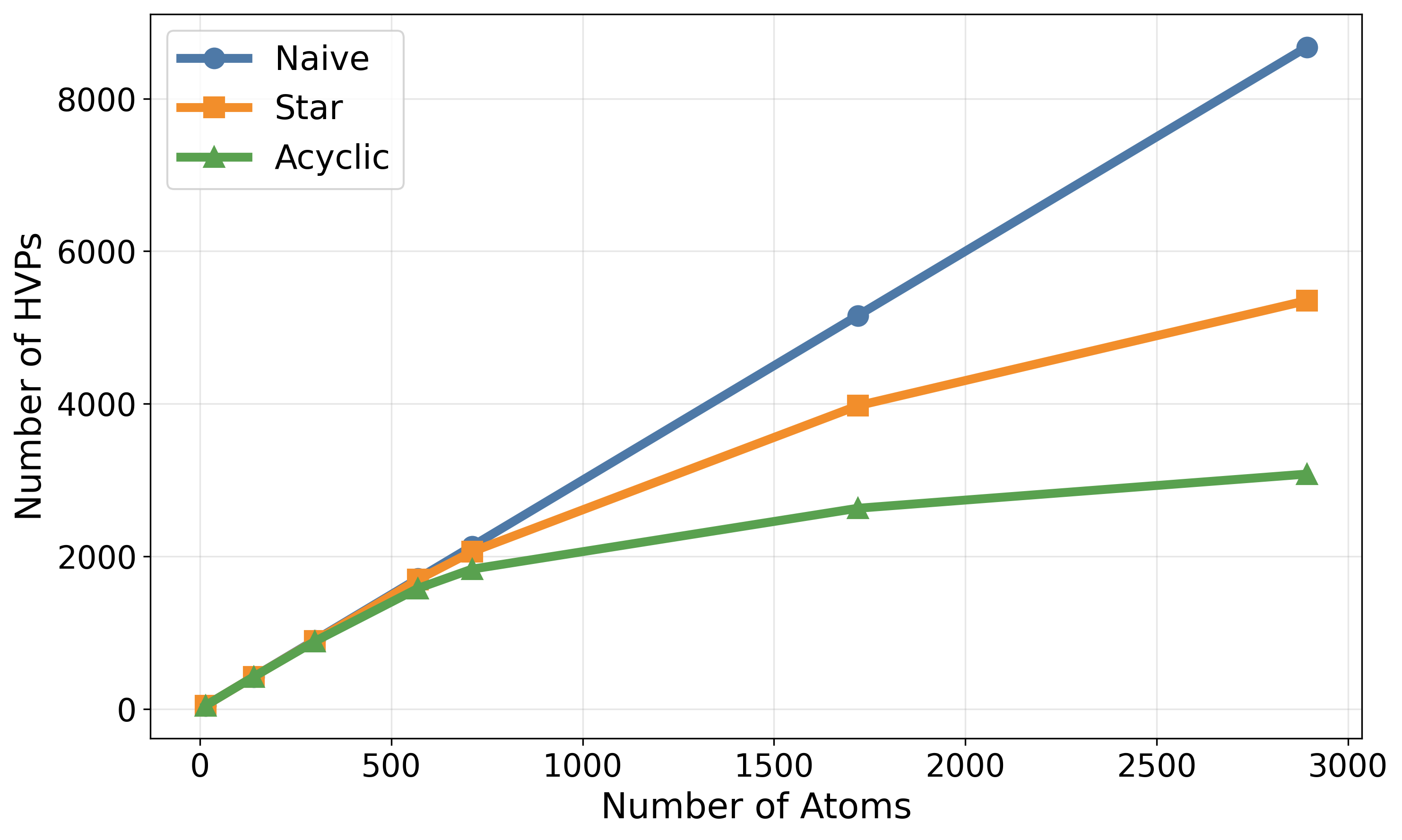}
    \caption{Water cluster HVP count scaling.}
\end{subfigure}
\caption{
The runtime and HVP count of a 2 Layer TensorNet for increasingly larger alkane chains and water clusters. \label{fig:alkanes_and_water_1l_extended}
}
\end{figure}

\FloatBarrier
\subsection{Protein Conformer}
\begin{figure}[ht]
\centering
    \centering
    \includegraphics[width=0.9\linewidth]{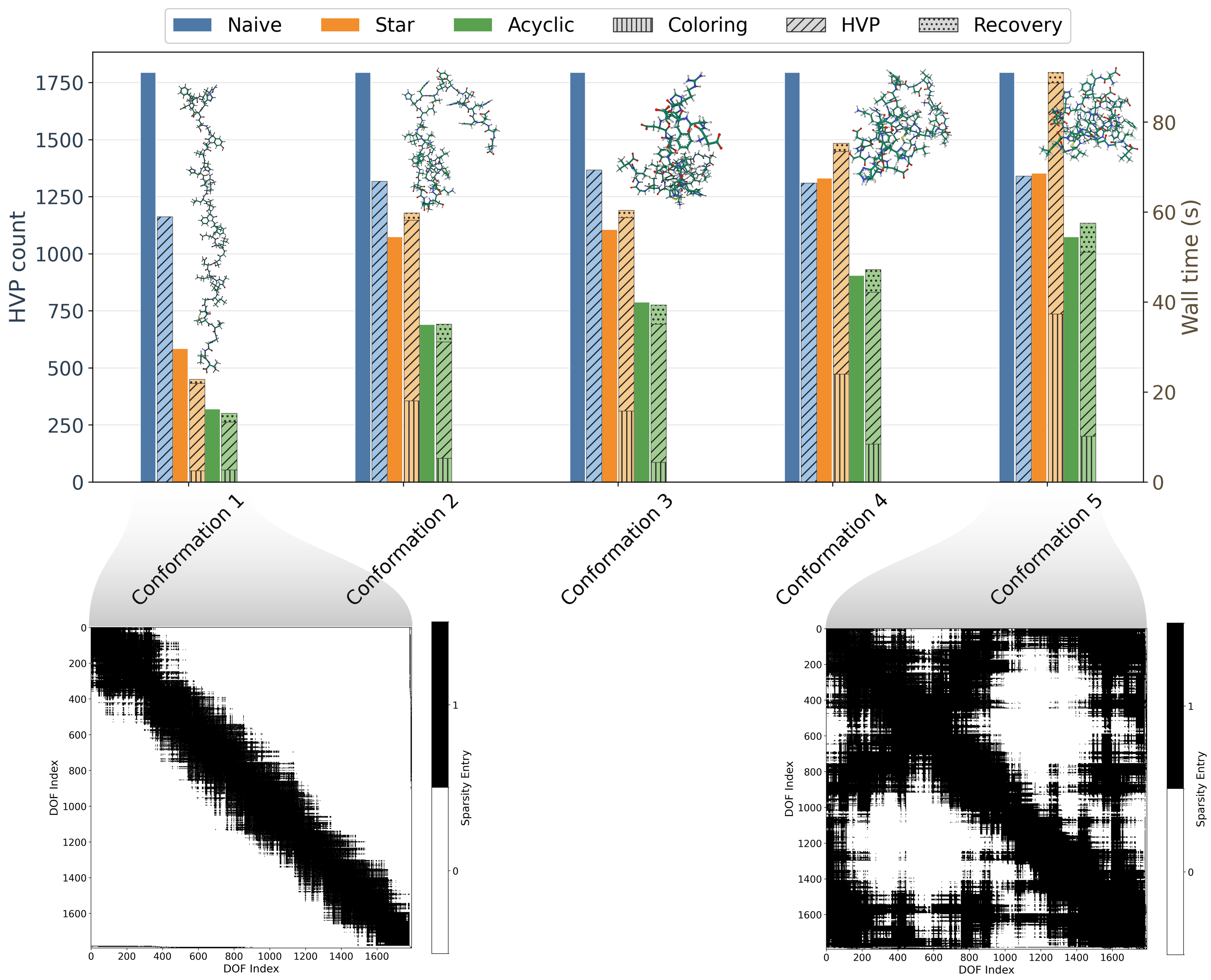}
\caption{Conformation-dependent sparse Hessian performance for five A$\beta$40 structures. Variation across conformations arises from differences in the induced graph sparsity pattern.\label{fig:abeta40_two_layers}}
\end{figure}

\FloatBarrier

\subsection{Cutoff experiments}
\label{sec:cutoff_details}
To use our sparse differentiation in conjunction with DFT and non-local MLIP Hessians, we have to use an approximate sparsity pattern based on atom pair distance cutoffs. For our DFT experiments, we are generating the Hessian with ORCA \cite{ORCA}. We use the PBE/sto-3g level of theory with the tight convergence criterion. Since the Chignolin geometry is far from equilibrium, convergence of the SCF is tricky, requiring second-order SCF and multiple restarts even in the sto-3g basis set. We used ORCA's tight convergence settings with a total compute time of 3 hours. Since ORCA is closed source and doesn't expose HVPs in their API, we emulate them using the full Hessian. In principle, CPKS allows for HVPs, though. Alternatively, we can implement HVPs using the finite difference approximation \ref{eq:finite_difference}. 

\begin{figure}[ht]
\centering
\begin{subfigure}[t]{0.49\linewidth}
    \centering
    \includegraphics[width=\linewidth]{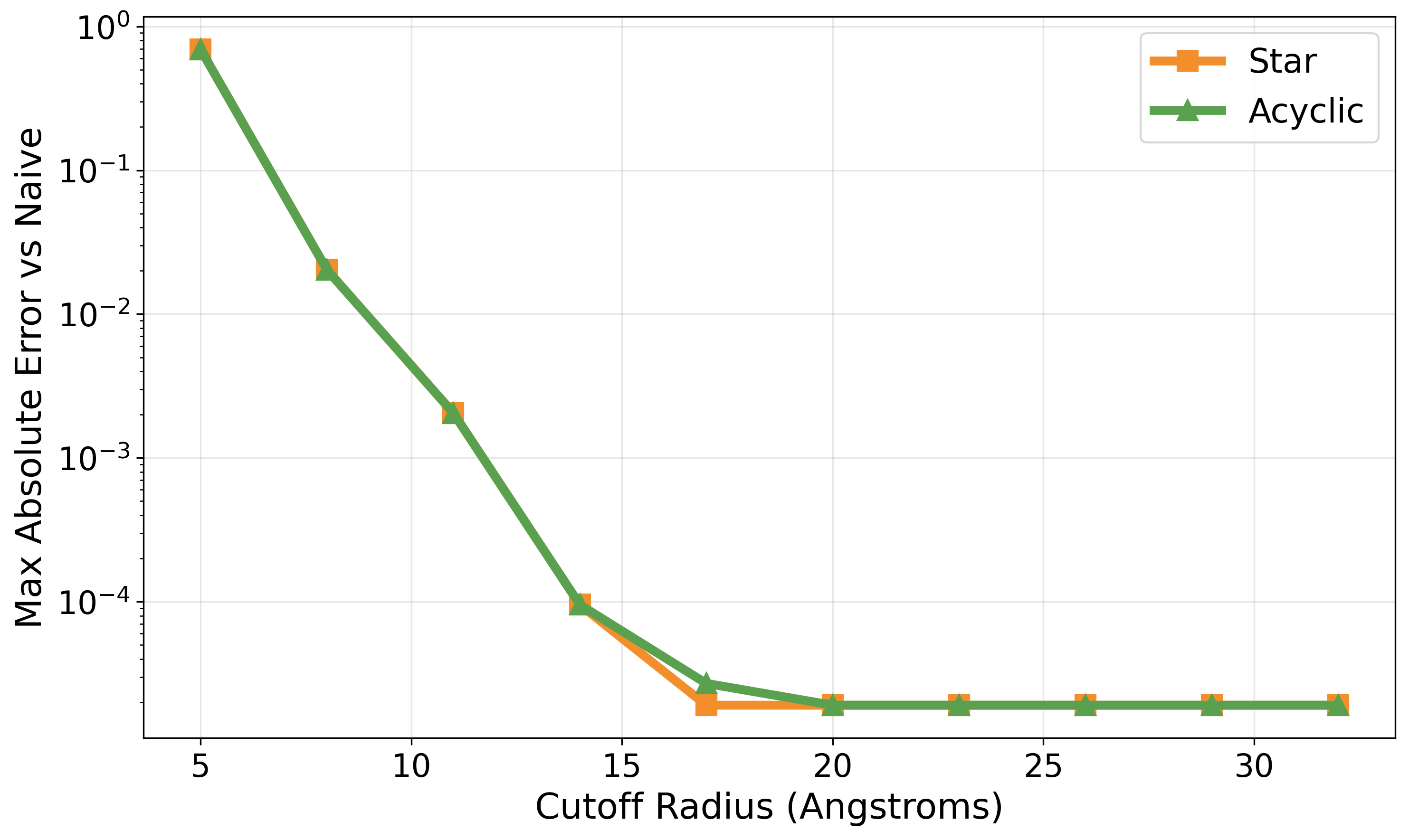}
    \caption{Maximum error vs cutoff radius on Chignolin with the large receptive field CHGNet model. \label{fig:cutoff_max_error_a}}
\end{subfigure}\hfill
\begin{subfigure}[t]{0.49\linewidth}
    \centering
    \includegraphics[width=\linewidth]{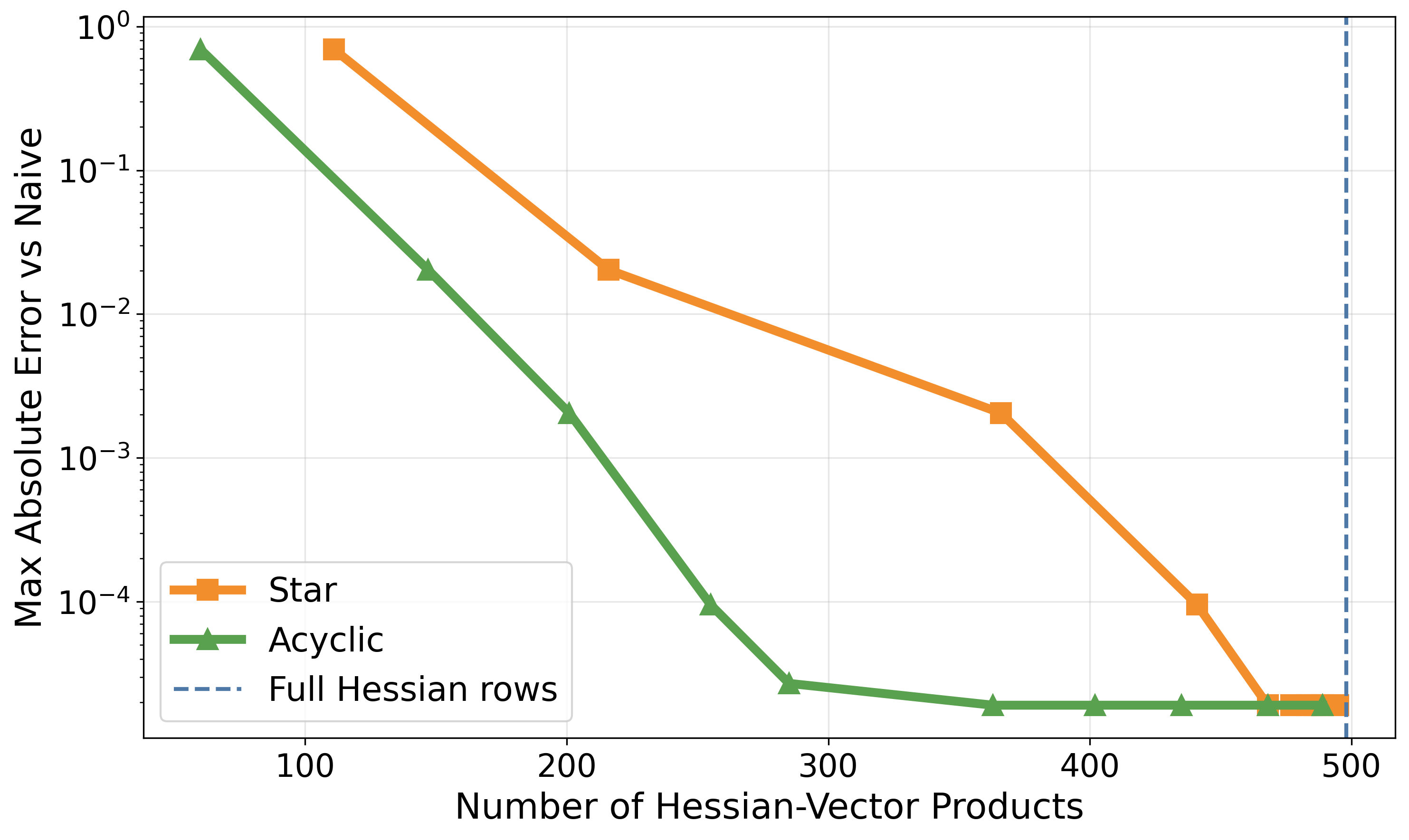}
    \caption{ Maximum error vs HVP count on Chignolin with the large receptive field CHGNet model. \label{fig:cutoff_max_error_b}}
\end{subfigure}
\begin{subfigure}[t]{0.49\linewidth}
    \centering
    \includegraphics[width=\linewidth]{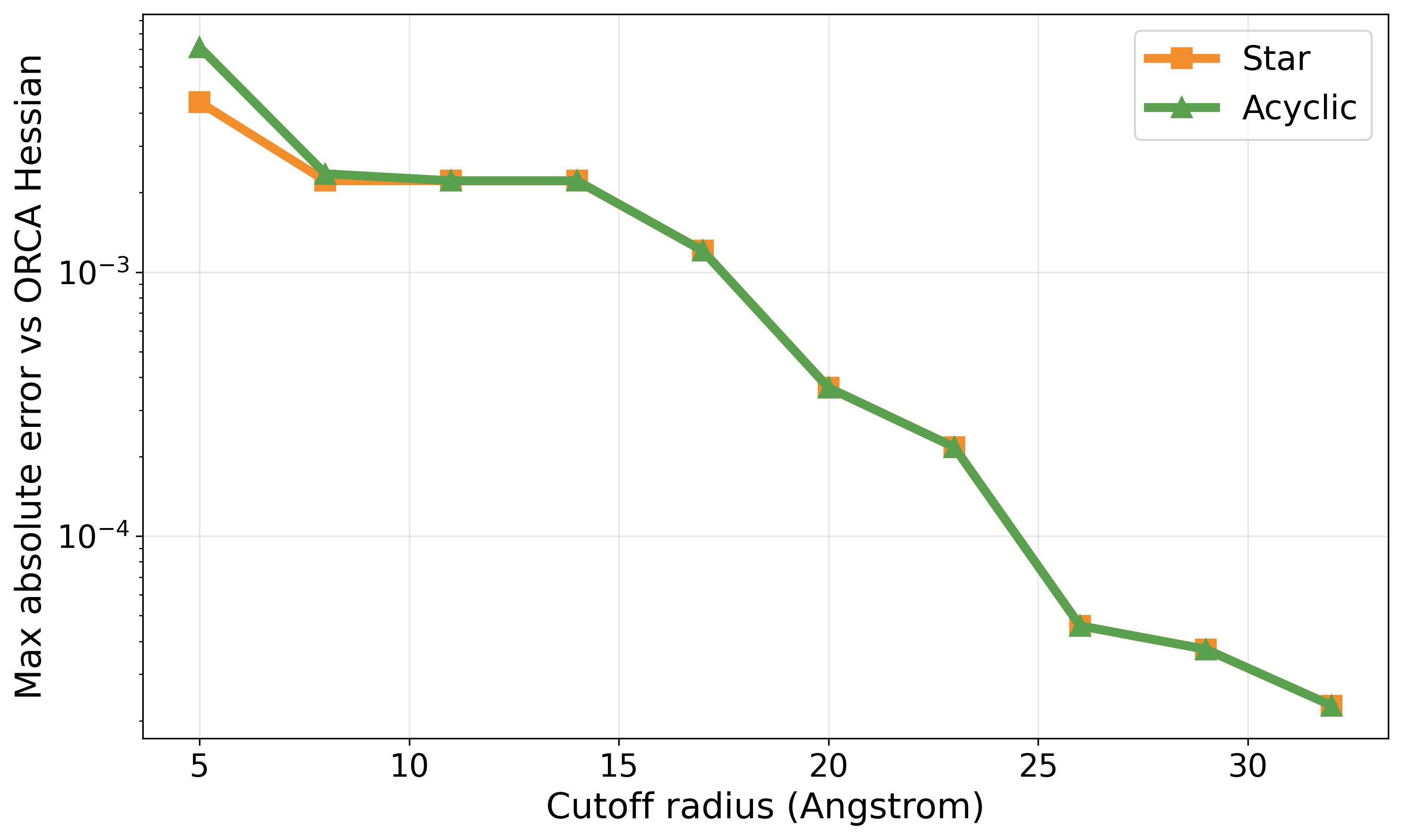}
    \caption{Maximum error vs cutoff radius on Chignolin with DFT Hessians. \label{fig:cutoff_max_error_c}}
\end{subfigure}\hfill
\begin{subfigure}[t]{0.49\linewidth}
    \centering
    \includegraphics[width=\linewidth]{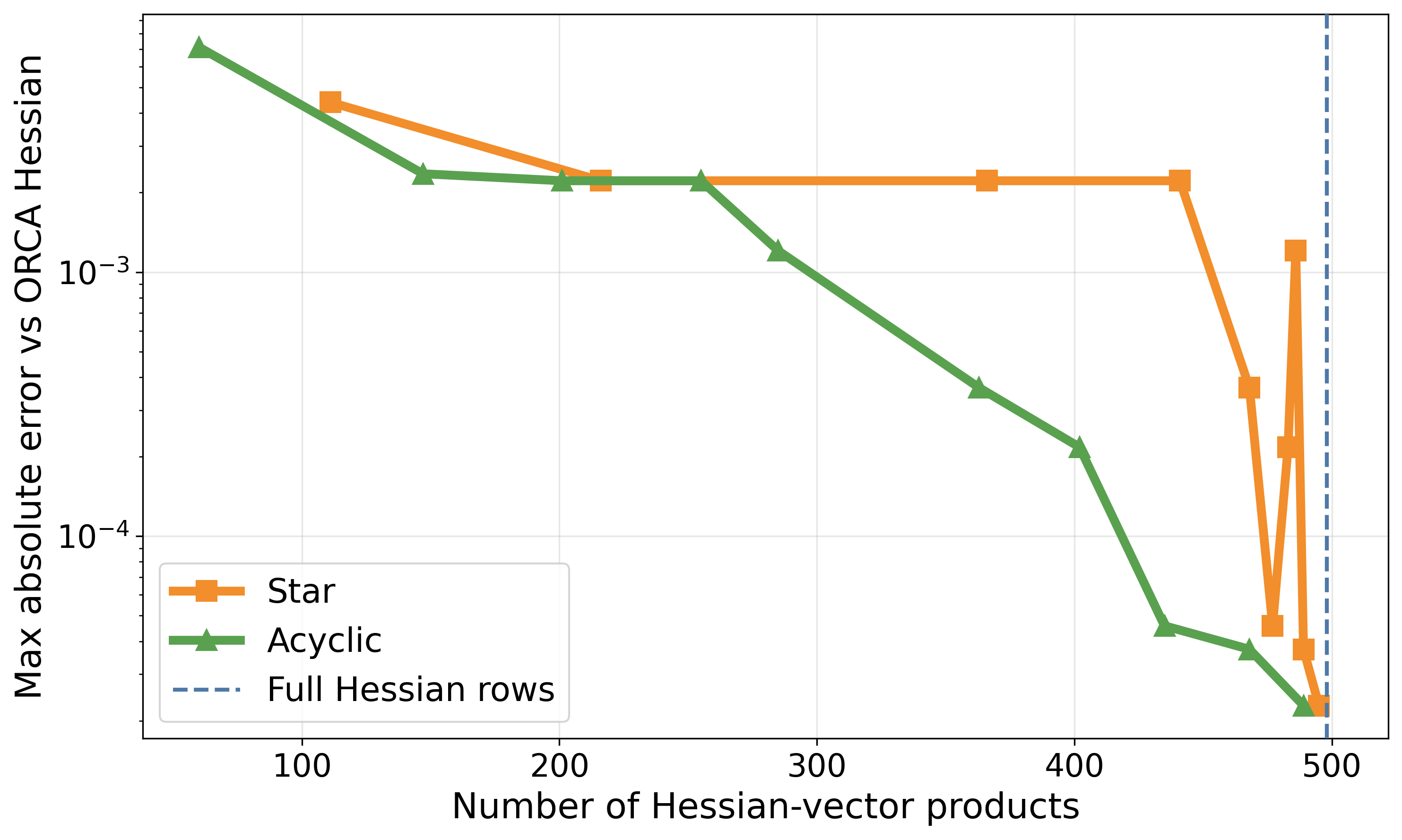}
    \caption{Maximum error vs HVP count on Chignolin with DFT Hessians. \label{fig:cutoff_max_error_d}}
\end{subfigure}
\caption{The maximum Hessian entry error of the CHGNet model (top) and DFT (bottom) resulting from using an approximate sparsity pattern.}

\end{figure}

Additionally to the average relative eigenvalue spectrum error from figure \ref{fig:cutoff_a} - \ref{fig:cutoff_d} we show maximal Hessian errors $\text{max}_{ij} |H_{ij}^\text{sparse} - H_{ij}^\text{naive}|$ in figure \ref{fig:cutoff_max_error_a} to \ref{fig:cutoff_max_error_d}.

\section*{Supporting Information}
The code is available at \url{https://github.com/the-matter-lab/sparse_hessian_ff}. 
The code includes our Python bindings for the ColPack C++ library \cite{gebremedhin2013colpack} and the integration with the Materials Graph Library (MatGL) \cite{ko2025matgl}, as well as the xyz files for the benchmarked systems.


\newpage

\end{document}